\documentclass[10pt,journal,compsoc]{IEEEtran}
\ifCLASSOPTIONcompsoc
  \usepackage[nocompress]{cite}
\else
  \usepackage{cite}
\fi
\ifCLASSINFOpdf
  \usepackage[pdftex]{graphicx}
\else
\fi
\newtheorem{definition}{Definition}
\newtheorem{lemma}{Lemma}
\newtheorem{theorem}{Theorem}

\usepackage{hyperref}
\usepackage{algorithm}
\usepackage{algorithmic}
\usepackage{enumitem}
\usepackage{xcolor}
\usepackage{amsmath}
\usepackage{amssymb}
\usepackage{fixmath}
\usepackage[switch, pagewise, displaymath, mathlines]{lineno}
\begin{document}
%
\title{Blind GB-PANDAS: A Blind Throughput-Optimal Load Balancing Algorithm for Affinity Scheduling}
%
%
%
%

\author{Ali Yekkehkhany, 
        and Rakesh Nagi
}

%
%


\markboth{IEEE/ACM TRANSACTIONS ON NETWORKING}%
{}

%



\IEEEtitleabstractindextext{%
\begin{abstract}
Dynamic affinity load balancing of multi-type tasks on multi-skilled servers, when the service rate of each task type on each of the servers is known and can possibly be different from each other, is an open problem for over three decades. The goal is to do task assignment on servers in a real time manner so that the system becomes stable, which means that the queue lengths do not 
diverge to infinity in steady state (throughput optimality), and the mean task completion time is minimized (delay optimality). The fluid model planning, Max-Weight, and c-$\mu$-rule algorithms have theoretical guarantees on optimality in some aspects for the affinity problem, but they consider a complicated queueing structure and either require the task arrival rates, the service rates of tasks on servers, or both. In many cases that are discussed in the introduction section, both task arrival rates and service rates of different task types on different servers are unknown. In this work, the Blind GB-PANDAS algorithm is proposed which is completely blind to task arrival rates and service rates. Blind GB-PANDAS uses an exploration-exploitation approach for load balancing. We prove that Blind GB-PANDAS is throughput optimal under arbitrary and unknown distributions for service times of different task types on different servers and unknown task arrival rates. Blind GB-PANDAS desires to route an incoming task to the server with the minimum weighted-workload, but since the service rates are unknown, such routing of incoming tasks is not guaranteed which makes the throughput optimality analysis more complicated than the case where service rates are known. Our extensive experimental results reveal that Blind GB-PANDAS significantly outperforms existing methods in terms of mean task completion time at high loads.
\end{abstract}

}

\maketitle

\IEEEdisplaynontitleabstractindextext

%
\IEEEpeerreviewmaketitle

\IEEEraisesectionheading{\section{Introduction} \label{introduction}}

Affinity load balancing refers to allocation of computing tasks on computing nodes in an efficient way to minimize a cost function, for example the mean task completion time \cite{padua2011encyclopedia}. Due to the fact that different task types can have different processing (service) rates on different computing nodes (servers), a dilemma between throughput and delay optimality emerges which makes the optimal affinity load balancing an open problem for more than three decades if the task arrival rates are unknown.
If the task arrival rates and the service rates of different task types on different servers are known, the fluid model planning algorithm by Harrison and Lopez \cite{harrison1998heavy, harrison1999heavy}, and Bell and Williams \cite{bell2001dynamic, bell2005dynamic} is a delay optimal load balancing algorithm that solves a linear programming optimization problem to determine task assignment on servers.
The same number of queues as the number of task types is needed for the fluid model planning algorithm, so the queueing structure is fixed to the number of task types and does not capture the complexity of the system model, which is how heterogeneous the service rates of task types on different servers are. As an example given in \cite{xie2016scheduling} and \cite{yekkehkhany2017near}, for data centers with a rack structure that use Hadoop for map-reduce data placement with three replicas of data chunks on the $M$ severs, the fluid model planning algorithm requires $M \choose 3$ queues, while Xie et al. \cite{xie2016scheduling} propose a delay optimal algorithm that uses $3M$ queues.
 As another extreme example, if the service rates of $N_T$ number of task types on all servers are the same, the fluid model planning algorithm still considers $N_T$ number of queues, while the First-Come-First-Served (FCFS) algorithm uses a single queue and is both throughput and delay optimal. It is true that in the last example all task types can be considered the same type, but this is just an example to enlighten the reasoning behind the queueing structure for GB-PANDAS (Generalized Balanced Priority Algorithm for Near Data Scheduling) presented in Section \ref{Queueing_Structure_for_GB-PANDAS}.

\begin{figure*}[t]
\centering
\includegraphics[scale=0.55]{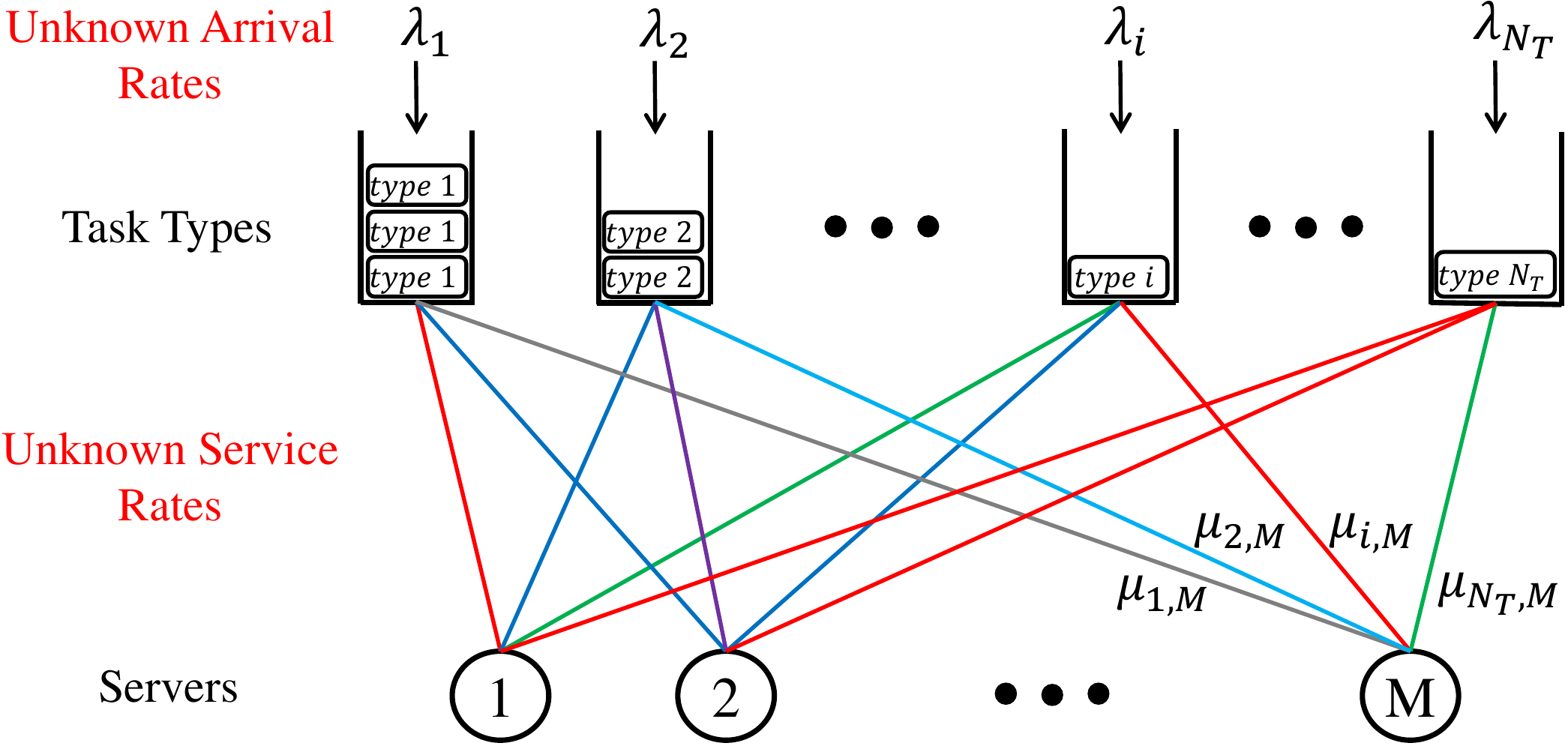}
\caption{Affinity scheduling setup with multi-type tasks and multi-skilled servers.}
\label{systemmodel}
\end{figure*}

In the absence of knowledge on task arrival rates, Max-Weight \cite{stolyar2004maxweight} and c-$\mu$-rule \cite{mandelbaum2004scheduling} algorithms can stabilize the system by just knowing the service rates of task types on different servers. None of these two algorithms are delay optimal though. The c-$\mu$-rule is actually cost optimum, where it assumes convex delay costs associated to each task type, and minimizes the total cost incurred to the system. Since the cost functions have to be strictly convex, so cannot be linear, c-$\mu$-rule does not minimize the mean task completion time. Since these two algorithms do not use the task arrival rates and still stabilize the system, they are robust to any changes in task arrival rate as long as it is in the capacity region of the system. Both Max-Weight and c-$\mu$-rule algorithms have the same issue as the fluid model planning algorithm on considering one queue per task type which can make the system model complicated as discussed in \cite{xie2016scheduling}. Note that Wang et al. \cite{wang2016maptask} and Xie et al. \cite{xie2016scheduling} study the load balancing problem for special cases of two and three levels of data locality, respectively. In the former, delay optimality is analyzed for a special traffic scenario and in the latter delay optimality is analyzed for a general traffic scenario and in both cases there is no issue on the number of queues, but as mentioned, these two algorithms are for special cases of two and three levels of data locality. Hence, a unified algorithm that captures the trade-off between the complexity of the queueing structure and the complexity of the system model is missed in the literature. Yekkehkhany et al. \cite{yekkehkhany2017gb} implicitly mention this trade-off in data center applications, but the generalization is not crystal clear and needs more thinking for the affinity setup, which is summarized in this work as a complementary note on the Balanced-PANDAS algorithm.


The affinity scheduling problem appears in different applications from data centers and modern processing networks that consist of heterogeneous servers, where data-intensive analytics like MapReduce, Hadoop, and Dryad are performed, to supermarket models, or even patient assignment to surgeons  in big and busy hospitals and many more. Lack of dependable estimates of system parameters, including task arrival rates and specially service rates of task types on different servers, is a major challenge in constructing an optimal load balancing algorithm for such networks \cite{pedarsani2017robust}. All the algorithms mentioned above at least require the knowledge of service rates of task types on different servers. In the absence of prior knowledge on service rates, such algorithms can be fragile and perform poorly, resulting in huge waste of resources. To address this issue, we propose a robust policy called Blind GB-PANDAS that is totally blind to all system parameters, but is robust to task arrival rate changes, learns the service rates of task types on different servers, so it is robust to any service rate parameter changes as well. It is natural that due to traffic load changes in data centers, the service rate of tasks on remote servers change over time. In such cases, Blind GB-PANDAS is capable of updating system parameters and taking action correspondingly. Blind GB-PANDAS uses an exploration-exploitation approach to make the system stable without any knowledge about the task arrival rates and the processing rates. More specifically, it uses an exploration-exploitation method, where in the exploration phase it takes action in a way to make the system parameter estimations more accurate, and in the exploitation phase it uses the estimated parameters to do an optimal load balancing based on the estimates. Note that only the processing rates of task types on different servers are the parameters that are estimated, and the task arrival rates are not estimated. The reason is that task arrival rates change frequently, so there is not a point on estimating them, whereas the service rates do not change rapidly.
Since Blind GB-PANDAS uses an estimate of the processing rates, an incoming task is not necessarily routed to the server with the minimum weighted-workload in the exploitation phase, which increases complexity in the throughput optimality proof of Blind GB-PANDAS using the Lyaponuv-based method.
The throughput optimality result is proved under arbitrary and unknown service time
 distributions with bounded means and bounded supports that do not necessarily require the memory-less property.

As discussed in Section \ref{Queueing_Structure_for_GB-PANDAS}, the queueing structure used for Blind GB-PANDAS shows the trade-off between the heterogeneity of the underlying system model for processing rates and the complexity of the Blind GB-PANDAS queueing structure. Blind GB-PANDAS can also use a one queue per server queueing structure, where the workload on servers is of interest instead of the queue lengths, but for an easier explanation of the Blind GB-PANDAS algorithm we use multiple symbolic sub-queues for each server. The Blind GB-PANDAS algorithm is compared to FCFS, Max-Weight, and c-$\mu$-rule algorithms in terms of average task completion time through simulations, where the same exploration-exploitation approach as Blind GB-PANDAS is used for Max-Weight and c-$\mu$-rule.
Our extensive simulations show that the Blind GB-PANDAS algorithm outperforms the three other algorithms at high loads by a reasonably large difference.

The rest of the paper is structured as follows. Section \ref{SystemModelSection} describes the system model, GB-PANDAS, and the queueing structure of GB-PANDAS, in addition to deriving the capacity region of the system. Section \ref{BP} presents the Blind GB-PANDAS algorithm and queueing dynamics for this algorithm. Section \ref{ThroughputOptimalitySection} starts with some preliminary results and lemmas and ends up with the throughput optimality proof for Blind GB-PANDAS. Section \ref{SimulationResultsSection} evaluates the performance of Blind GB-PANDAS versus Max-Weight, c-$\mu$-rule, and FCFS algorithms in terms of mean task completion time. Section \ref{RelatedWorkSection} discusses the related works, and finally Section \ref{ConclutionSection} concludes the paper with a discussion on opportunities for future work.

\section{System Model}
\label{SystemModelSection}

\begin{figure}[t]
\centering
\includegraphics[scale=0.49]{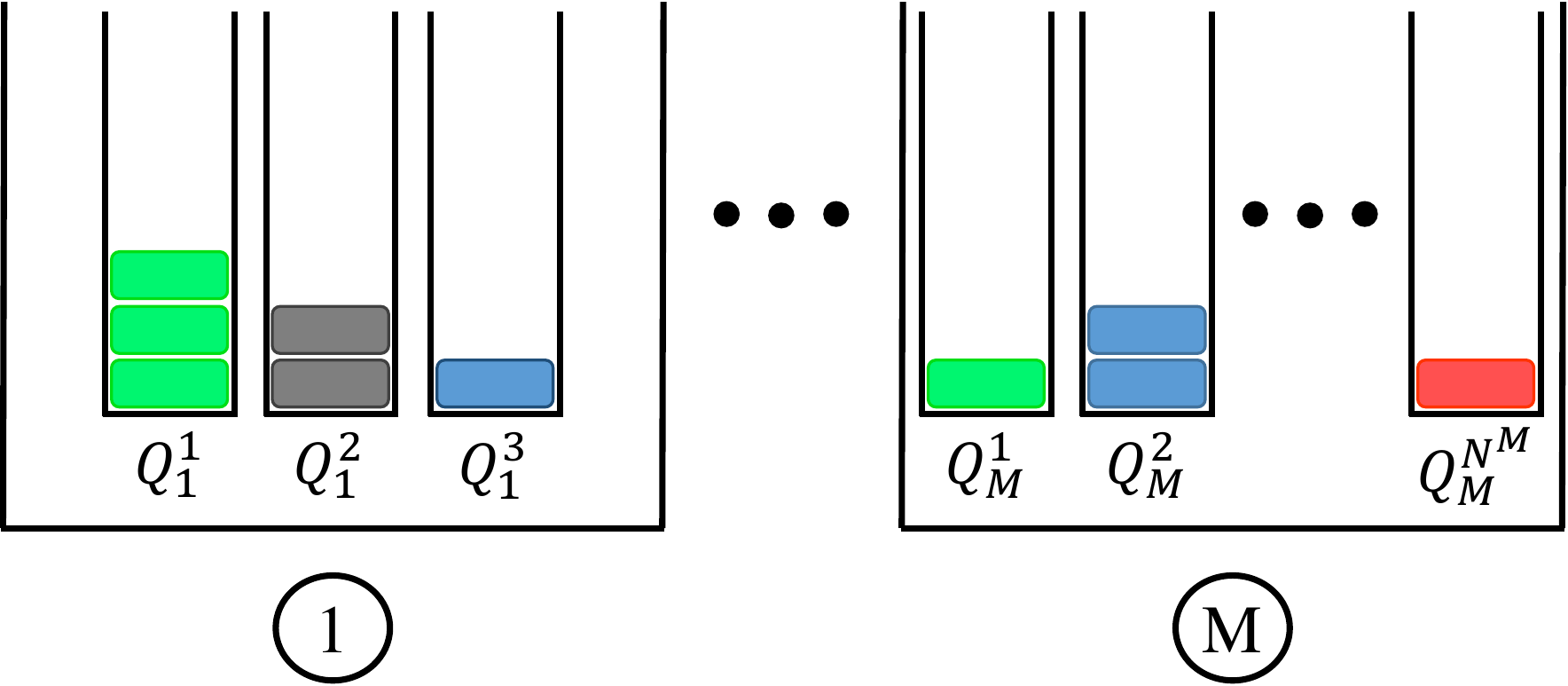}
\caption{The queueing structure for the GB-PANDAS algorithm.}
\label{queueing_structure}
\end{figure}

Consider $M$ unit-rate multi-skilled servers and $N_T$ number of task types as depicted in Figure \ref{systemmodel}. The set of servers and task types are denoted by $\mathcal{M} = \{1, 2, \cdots, M\}$ and $\mathcal{L} = \{1, 2, \cdots, N_T\},$ respectively. Each task can be processed by any of the $M$ servers, but with possibly different rates.
The service times are assumed to be non-preemptive and discrete valued with an unknown distribution. Non-preemptive service means that the central load balancing algorithm cannot interrupt an in-service task, i.e. no other task is scheduled to a server until the server completely processes the task that is currently receiving service. The extension of the analysis for continuous service time, using approximation methods of continuous distributions with discrete ones, is an interesting future work.
 In this discrete time model, time is indexed by $t \in \mathbb{N}$. In the following, service time distributions and task arrivals are discussed, which are both unknown to the central scheduler. \\
\textbf{Service time distribution:} The service time offered by server $m \in \mathcal{M}$ to task type $i \in \mathcal{L}$ is a discrete-type random variable with cumulative distribution function (CDF) $F_{i, m}$ with mean $\frac{1}{\mu_{i, m}}$ or correspondingly with rate $\mu_{i, m} > 0$. The service time distribution does not require the memory-less property. We further assume that the support of the service time is bounded,
which is a realistic assumption and reduces the unnecessary complexity of the proofs specially in Lemma \ref{lemma1234}.
The extension of the analysis for service times with unbounded supports is an interesting future work. Note that the completion time for a task is the waiting time for that task until it is scheduled to a server plus the service time of the task on the server. Waiting time depends on the servers' status, the queue lengths or more specifically other tasks that are in the system or may arrive later, and the load balancing algorithm that is used, while service time has the mentioned distribution. \\
\textbf{Task arrival:} The number of incoming tasks of type $i \in \mathcal{L}$ at the beginning of time slot $t$ is a random variable on non-negative integer numbers that is denoted by $A_i(t)$, which are temporarily identically distributed and independent from each other. Denote the arrival rate of task type $i$ by $\lambda_i$, i.e. $\mathbb{E}[A_i(t)] = \lambda_i$. In the stability proof of Blind GB-PANDAS we need $\lambda_i$ to be strictly positive, so without loss of generality we exclude task types with zero arrival rate from $\mathcal{L}$. Furthermore, we assume that the number of each incoming task type at a time slot is bounded by constant $C_A$ and is zero with positive probability, i.e. $P(A_i(t) < C_A) = 1$ and $P(A_i(t) = 0) > 0$ for any $i \in \mathcal{L}$. The set of arrival rates for all task types is denoted by vector $\mathbold{\lambda} = (\lambda_i: i \in \mathcal{L})$.

Affinity scheduling problem refers to load balancing for such a system described above. The fluid model planning algorithm \cite{harrison1999heavy}, MaxWeight \cite{stolyar2004maxweight}, and c$\mu$-rule \cite{mandelbaum2004scheduling} are the baseline algorithms for affinity scheduling. 
All these algorithms in addition to GB-PANDAS use the rate of service times instead of the CDF functions.  Hence, the system model can be summarized as an $N_T \times M$ matrix, where element $(i, m)$ is the processing rate of task type $i$ on server $m$, $\mu_{i, m}$, as follows:
\begin{equation}
B_\mu = \begin{bmatrix}
    \mu_{1,1} & \mu_{1,2} & \mu_{1,3} & \dots  & \mu_{1, M} \\
    \mu_{2,1} & \mu_{2,2} & \mu_{2,3} & \dots  & \mu_{2, M} \\
    \vdots & \vdots & \vdots & \ddots & \vdots \\
    \mu_{N_T,1} & \mu_{N_T,2} & \mu_{N_T,3} & \dots  & \mu_{N_T, M}
\end{bmatrix}_{N_T, M}.
\end{equation}
If both the set of arrival rates $\mathbold{\lambda} = (\lambda_i : i \in \mathcal{L})$ and the service rate matrix $B_\mu$ are known, the fluid model planning algorithm \cite{harrison1999heavy} derives the delay optimal load balancing by solving a linear programming. However, if the arrival rates of task types are not known, the delay optimal algorithm becomes an open problem which has not been solved for more than three decades. Max-Weight \cite{stolyar2004maxweight} and c$\mu$-rule \cite{mandelbaum2004scheduling} can be used for different objectives when we do not know the arrival rates, but none have delay optimality. In this work, we are assuming that we lack knowledge of not only the arrival rates $\mathbold{\lambda}$, but also the service rate matrix $B_\mu$. We take an exploration and exploitation approach to make our estimation of the underlying model, which is the service rate matrix, more accurate, and to keep the system stable.


\subsection{Queueing Structure for GB-PANDAS}
\label{Queueing_Structure_for_GB-PANDAS}

Every algorithm has its own specific queueing structure. For example, there is only a single central queue for the First-Come-First-Served (FCFS) algorithm, but there are $N_T$ number of queues when using fluid model planning, Max-Weight, or c$\mu$-rule. In the following, we present the queueing structure used for GB-PANDAS that captures the trade-off between the complexity of the system model and the complexity of the queueing structure very well. 
 What we mean by the complexity of the system model is the heterogeneity of the service rate matrix, e.g. if all the elements of this matrix are the same number, the system is less complex than the case where each element of the matrix is different from other elements of the matrix.

The heterogeneity of the system from the perspective of server $m$ is captured in the $m$\textsuperscript{th} column of the service rate matrix. Consider the $m$\textsuperscript{th} column of the matrix has $N^m$ distinct values, where $N^m$ can be any number from $1$ to $N_T$. It is obvious that any of the task types with the same service (processing) rate on server $m$ look the same from the perspective of this server. Denote the $N^m$ distinct values of the $m$\textsuperscript{th} column of $B_\mu$ by $\{\alpha_m^1, \alpha_m^2, \cdots, \alpha_m^{N^m}\}$ and without loss of generality assume that $\alpha_m^1 > \alpha_m^2 > \cdots > \alpha_m^{N^m}$. We call all the task types with a processing rate of $\alpha_m^n$ on the $m$\textsuperscript{th} server, the $n$-local tasks to that server, and denote them by $\mathcal{L}_m^n = \{ i \in \mathcal{L}: \mu_{i, m} = \alpha_m^n \}$. For ease of notation, we use both $\mu_{i, m}$ and $\alpha_m^n$ throughout the paper interchangeably; however, they are in fact capturing the same phenomenon, but with different interpretations. Note that the $n$-local tasks to server $m$ can be called $(n,m)$-local tasks in order to place more emphasis on the pair $n$ and $m$, so the $n$-local tasks to server $m$ are not necessarily the same as the $n$-local tasks to server $m'$. We allocate $N^m$ queues for server $m$, where the $n$\textsuperscript{th} queue of server $m$ holds all task types that are routed to this server and are $n$-local to it.
As depicted in Figure \ref{queueing_structure}, different servers can have different numbers of queues since the heterogeneity of the system model can be different from the perspective of different servers. We may interchangeably use queue or sub-queue to refer to the $n$\textsuperscript{th} queue (sub-queue) of the $m$\textsuperscript{th} server. The $N^m$ sub-queues of the $m$\textsuperscript{th} server are denoted by $Q_m^1, Q_m^2, \cdots, Q_m^{N^m}$ and the queue lengths of these sub-queues, defined as the number of tasks in these sub-queues, at time slot $t$ are denoted by $Q_m^1(t), Q_m^2(t), \cdots, Q_m^{N^m}(t)$.

In the next subsection, the GB-PANDAS algorithm is proposed when the service rate matrix $B_\mu$ is known.
Balanced-PANDAS for a data center with three levels of data locality is proposed by \cite{xie2016scheduling}, and here we are proposing the Generalized Balanced-PANDAS algorithm from another perspective which is of its own interest. 


\subsection{GB-PANDAS Algorithm with Known Service Rate Matrix $B_{\mu}$}
\label{GB-PANDAS_K}
Before getting into the GB-PANDAS algorithm, we need to define the workload on server $m$.
\begin{definition}
The average time needed for server $m$ to process all tasks queued in its $N^m$ sub-queues at time slot $t$ is defined as the workload on the server:
\begin{equation}
\label{exactWW}
W_m(t) = \frac{Q_m^1(t)}{\alpha_m^1} + \frac{Q_m^2(t)}{\alpha_m^2} + \cdots + \frac{Q_m^{N^m}(t)}{\alpha_m^{N^m}}.
\end{equation}
\end{definition}

A load balancing algorithm consists of two parts, routing and scheduling. The routing policy determines the queue at which an incoming task is stored until it is assigned to a server for service. When a server becomes idle, the scheduling policy determines the next task that receives service on the idle server. The routing and scheduling policies of the GB-PANDAS algorithm are as follows: \\
\textbf{GB-PANDAS Routing Policy:} An incoming task of type $i$ is routed to the corresponding sub-queue of the server with the minimum weighted workload, where ties are broken arbitrarily to the favor of the fastest server. The server $m^*$ with the minimum weighted workload is defined as
\[
m^* = \underset{m \in \mathcal{M}}{\arg \min} \ \frac{W_m(t)}{\mu_{i, m}}.
\]
The corresponding sub-queue of server $m^*$ for a task of type $i$ is $n$ if $\mu_{i, m} = \alpha_m^n$. \\
\textbf{GB-PANDAS Scheduling Policy:} An idle server $m$ at time slot $t$ is scheduled to process a task of sub-queue $Q_m^1$ if there is any. If $Q_m^1(t) = 0,$ a task of sub-queue $Q_m^2$ is scheduled to the server, and so on. It is a common assumption that servers do not have the option of processing the tasks queued in front of other servers, so a server remains idle if all its sub-queues are empty. Note that the routing policy is doing a sort of weighted water-filling for workloads, so the probability that a server becomes idle goes to zero as the load increases at heavy traffic regime.
Remember that the tasks in sub-queue $Q_m^1$ are the fastest types of tasks for server $m$, the tasks in sub-queue $Q_m^2$ are the second fastest, and so on. Using this priority scheduling, the faster tasks in the $N^m$ sub-queues of server $m$ are processed first. Given the minimum weighted workload routing policy, the priority scheduling is optimal as it minimizes the mean task completion time of all tasks in the $N^m$ sub-queues of server $m$.
In the following, Max-Weight and c$\mu$-rule algorithms are discussed for the sake of completeness.

\textbf{Remark.} Prioritized scheduling has no effect in the throughput-optimality proof of the GB-PANDAS algorithm and a work conservative scheduling of a server to its sub-queues suffices for the purpose of system's stability. As a result, the GB-PANDAS policy can be implemented by considering a single queue per server at the expense of losing priority scheduling.
In a single queue per server structure, instead of maintaining a server's sub-queue lengths, the workload of the server defined in \eqref{exactWW} is maintained. At the arrival of an $n$-local task to server $m$, the server's workload is increased by $\frac{1}{\alpha_m^n}$, instead of increasing the corresponding sub-queue's length by one, and the workload is decreased at the departure of a task by its corresponding load.

\subsection{Max-Weight and c-$\mu$-Rule Algorithms with Known Service Rate Matrix $B_{\mu}$}
\label{MWCMU}
The queueing structure used for Max-Weight and c-$\mu$-rule is as depicted in Figure \ref{systemmodel}, where there is a separate queue for each type of task. Denote the $N_T$ queues by $Q_1, Q_2, \cdots, Q_{N_T}$, and their corresponding queue lengths at time slot $t$ by $Q_1(t), Q_2(t), \cdots, Q_{N_T}(t)$. Note that the GB-PANDAS algorithm requires $M \times N_T$ number of queues in the worst case scenario, but it can use the symmetry of specific real-world structures to decrease the number of queues dramatically.
As an example, for servers with rack structures, where Hadoop is used for map-reduce data placement with three replicas of data chunks on severs, Max-Weight and c-$\mu$-rule require ${M \choose 3} = O(M^3)$ number of queues, while GB-PANDAS requires $3M$ queues.
A task is routed to a server at the time of its arrival under the GB-PANDAS algorithm, while a task waits in its queue under both Max-Weight and c-$\mu$-rule algorithms, waiting to be scheduled for service, which is discussed below. \\
\textbf{Max-Weight Scheduling Policy:} An idle server $m$ at time slot $t$ is scheduled to process a task of type $j$ from $Q_j$, if there is any, such that
\[
j \in \underset{i \in \mathcal{L}}{\arg\max} \ \{ \mu_{i, m} \cdot Q_i(t)  \}.
\]
The Max-Weight algorithm is throughput-optimal, but it is not heavy-traffic or delay optimal \cite{stolyar2004maxweight}. \\
\textbf{C-$\mu$-rule Scheduling Policy:} Consider that queue $Q_i$ incurs a cost of $C_i \big (Q_i(t) \big )$ at time slot $t$, where $C_i(.)$ is increasing and strictly convex. The c-$\mu$-rule algorithm maximizes the rate of decrease of the instantaneous cost at all time slots by the following scheduling policy. An idle server $m$ at time slot $t$ is scheduled to process a task of type $j$ from $Q_j$, if there is any, such that
\[
j \in \underset{i \in \mathcal{L}}{\arg\max} \ \big \{ \mu_{i, m} \cdot C_i' \big ( Q_i(t) \big ) \big \},
\]
where $C'(.)$ is the first derivative of the cost function. The c-$\mu$-rule algorithm minimizes both instantaneous and cumulative queueing costs, asymptotically. The mean task completion time corresponds to linear cost functions for all task types, so c-$\mu$-rule cannot minimize the mean task completion time, and as the result, is not heavy-traffic optimal.

\subsection{Capacity Region of Affinity Scheduling Setup}
\label{CapacityRegionSection}
We propose a decomposition of the arrival rate vector $\mathbold{\lambda} =$ $\left ( \lambda_i : i \in \mathcal{L} \right )$ as follows. For any task type $i \in \mathcal{L}, \lambda_i$ is decomposed into $\left ( \lambda_{i, m}, m \in \mathcal{M} \right )$, where $\lambda_{i, m}$ is assumed to be the arrival rate of type $i$ tasks for server $m$.
Hence, $\lambda_i = \sum_{m = 1}^M \lambda_{i, m}$.
By using the fluid model planning algorithm, the affinity queueing system can be stabilized under a given arrival rate vector $\mathbold{\lambda}$ as long as the necessary condition of total $1$-local, $2$-local, ...,  and $N^m$-local load on server $m$ being strictly less than one for any server $m$ is satisfied:
\begin{equation}
\label{necessarycondition}
\sum_{i \in \mathcal{L}} \frac{\lambda_{i, m}}{\mu_{i, m}} < 1, \ \ \ \ \ \forall m \in \mathcal{M}.
\end{equation}
Hence, the capacity region of the affinity problem is the set of all arrival rate vectors $\mathbold{\lambda}$ that has a decomposition $(\lambda_{i, m}, i \in \mathcal{L}, m \in \mathcal{M})$ satisfying \eqref{necessarycondition}:

\begin{equation}
\begin{aligned}
\label{capacityregion}
& \Lambda = \big \{ \boldsymbol{\lambda} = (\lambda_i : i \in \mathcal{L}) \ \big | \ \exists \lambda_{i, m} \geq 0, \forall i \in \mathcal{L}, \forall m \in \mathcal{M}, s.t. \\
& \ \ \ \ \ \ \ \ \ \ \lambda_i = \sum_{m = 1}^{M} \lambda_{i, m}, \ \forall i \in \mathcal{L}, \ \ \  \sum_{i \in \mathcal{L}} \frac{\lambda_{i, m}}{\mu_{i, m}} < 1, \forall m \in \mathcal{M} \big \}.
\end{aligned}
\end{equation}
The linear programming optimization associated with Equation \eqref{capacityregion} can be solved to find the capacity region $\Lambda$ of the system. The GB-PANDAS algorithm stabilizes the system for any arrival rate vector inside the capacity region by knowing the service rate matrix. It is proved in Section \ref{ThroughputOptimalitySection} that the Blind GB-PANDAS algorithm is throughput-optimal without the knowledge of the service rate matrix, $B_\mu$.

\section{The Blind GB-PANDAS Algorithm}
\label{BP}
The GB-PANDAS and Max-Weight algorithms need to know the precise value of the service rate matrix, but this requirement is not realistic for real applications. Furthermore, the service rate matrix can change over time, which confuses the load balancing algorithm if it uses a fixed given service rate matrix.
In the Blind version of GB-PANDAS, the service rate matrix is initiated randomly and is updated as the system is running. More specifically, an exploration-exploitation framework is combined with GB-PANDAS. In the exploration phase, the routing and scheduling are performed so as to allow room for making the estimations of the system parameters more precise, and in the exploitation phase the routing and scheduling are done based on the available estimation of the service rate matrix so as to stabilize the system.
Here we assume that $N^m$ is known as well as the locality level of a task on servers
that can be inferred from prior knowledge on the structure of the system. This is not a necessary assumption for throughput-optimality proof, but it makes the intuition behind Blind GB-PANDAS more clear.
As mentioned before, a single queue per server can be used when using Blind GB-PANDAS, in which case, there is no need to know $N^m$ as well as the ordering of service rates offered by servers for different task types.

\begin{figure*}[t]
\centering
\includegraphics[scale=0.55]{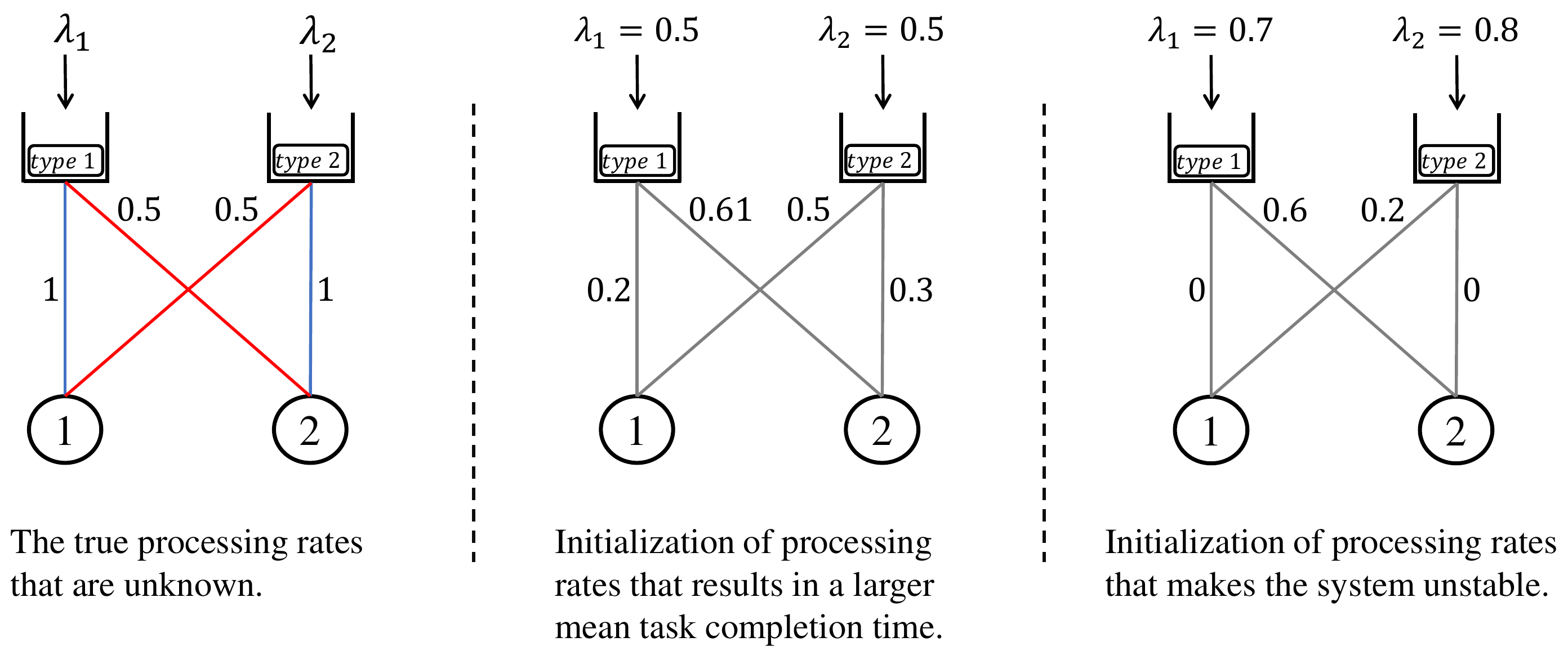}
\caption{This example shows that a queueing system with unknown processing rates can even be unstable for some initialization of processing rates if there is no exploration in the load balancing algorithm.}
\label{counter_example}
\end{figure*}

We first propose the updating method used for the service rate matrix before getting into the routing and scheduling policies of the Blind GB-PANDAS algorithm. The estimated service rate matrix at time slot $t$ is denoted as
\begin{equation}
\label{service_rate_matrix_estimation}
\hspace{-0.05cm} \widetilde{B}_\mu(t) = \begin{bmatrix}
    \widetilde{\mu}_{1,1}(t) & \widetilde{\mu}_{1,2}(t) & \widetilde{\mu}_{1,3}(t) & \dots  & \widetilde{\mu}_{1, M}(t) \\
    \widetilde{\mu}_{2,1}(t) & \widetilde{\mu}_{2,2}(t) & \widetilde{\mu}_{2,3}(t) & \dots  & \widetilde{\mu}_{2, M}(t) \\
    \vdots & \vdots & \vdots & \ddots & \vdots \\
    \widetilde{\mu}_{N_T,1}(t) & \widetilde{\mu}_{N_T,2}(t) & \widetilde{\mu}_{N_T,3}(t) & \dots  & \widetilde{\mu}_{N_T, M}(t)
\end{bmatrix}.
\end{equation}
Note that $\widetilde{\alpha}_m^1(t), \widetilde{\alpha}_m^2(t), \cdots, \widetilde{\alpha}_m^{N^m}(t), \ \forall m \in \mathcal{M}$ which are the estimates of $\alpha_m^1(t), \alpha_m^2(t), \cdots, \alpha_m^{N^m}(t), \ \forall m \in \mathcal{M}$ at time slot $t$ are nothing but the distinct values of the elements of the service rate matrix. More specifically, those are the $\widetilde{\alpha}_m^n, \ \forall m \in \mathcal{M}, \ \forall n \in \{1, 2, \cdots,$ $N^m\}$ that are getting updated and then mapped into their corresponding elements in the service rate matrix to form $\widetilde{B}_\mu$ in \eqref{service_rate_matrix_estimation} as mentioned in Section \ref{Queueing_Structure_for_GB-PANDAS}. Consider a random initialization of $\widetilde{\alpha}_m^n(0) > 0, \ \forall m \in \mathcal{M}, \ \forall n \in \{1, 2, \cdots, N^m\}$ at time slot $t = 0$. If server $m$ has processed $\widetilde{n} - 1$ tasks that are $n$-local to this server by time $t_1$, the estimate of $\alpha_m^n$ at this time slot is $\widetilde{\alpha}_m^n(t_1)$, and a new observation of service time for $n$-local task to server $m$ is made at time slot $t_2 > t_1$ as $T_m^n(t_2)$, we have $\widetilde{\alpha}_m^n(t) = \widetilde{\alpha}_m^n(t_1)$ for $t_1 \leq t < t_2$ and the update of this parameter at time slot $t_2$ is
\begin{equation}
\label{estimators}
\widetilde{\alpha}_m^n(t_2) = \frac{\widetilde{n} - 1}{\widetilde{n}} \cdot \widetilde{\alpha}_m^n(t_1) + \frac{1}{\widetilde{n} \cdot T_m^n(t_2)}.
\end{equation}
Note that $\widetilde{\alpha}_m^n$ is the service rate, not the service time mean, that is why $\frac{1}{T_m^n(t_2)}$ is used above in the update of the service rate. In the following, the routing and scheduling policies of Blind GB-PANDAS are presented, where the exploration rate is chosen in such a way that infinitely many $n$-local tasks are scheduled for service on server $m$ for any $m \in \mathcal{M}$ and any $n \in \{1, 2, \cdots, N^m\}$ so that by using the strong law of large numbers, the parameter estimations in \eqref{estimators} converge to their real values almost surely. \\
\textbf{Blind GB-PANDAS Routing Policy:}
The estimated workload on server $m$ at time slot $t$ is defined based on parameter estimations in \eqref{estimators} as
\begin{equation}
\label{estimatedWW}
\widetilde{W}_m(t) = \frac{Q_m^1(t)}{\widetilde{\alpha}_m^1(t)} + \frac{Q_m^2(t)}{\widetilde{\alpha}_m^2(t)} + \cdots + \frac{Q_m^{N^m}(t)}{\widetilde{\alpha}_m^{N^m}(t)}.
\end{equation}
The routing of an incoming task is based on the following exploitation policy with probability $p_e = max(1 - p(t), 0)$
, and is based on the exploration policy otherwise, where $p(t) \rightarrow 0$ as $t \rightarrow \infty$ and $\sum_{t = 0}^\infty p(t) = \infty$, e.g. the exploitation probability can be chosen as $p_e = 1 - \frac{1}{t^c}$ for $0 < c \leq 1.$
\begin{itemize}[leftmargin=*]
	\item \textbf{Exploitation phase:} An incoming task of type $i$ is routed to the corresponding sub-queue of the server with the minimum estimated weighted workload, where ties are broken arbitrarily. The server $\widetilde{m}^*$ with the minimum weighted workload for task of type $i$ is defined as
\[
\widetilde{m}^* = \underset{m \in \mathcal{M}}{\arg \min} \ \frac{\widetilde{W}_m(t)}{\widetilde{\mu}_{i, m}(t)}.
\]
The corresponding sub-queue of server $\widetilde{m}^*$ for a task of type $i$ is $n$ if $\widetilde{\mu}_{i, \widetilde{m}^*} = \widetilde{\alpha}_{\widetilde{m}^*}^n$.
	\item \textbf{Exploration phase:} An incoming task of type $i$ is routed to the corresponding sub-queue of a server chosen uniformly at random among $\{1, 2, \cdots, M\}$.
\end{itemize}
\textbf{Blind GB-PANDAS Scheduling Policy:}
The scheduling of an idle server is based on the following exploitation policy with probability $p_e$, and is based on the exploration policy otherwise.
\begin{itemize}[leftmargin=*]
	\item \textbf{Exploitation phase:} Priority scheduling is performed for an idle server as discussed in Section \ref{GB-PANDAS_K}. We emphasize that given the routing policy, priority scheduling is the optimal scheduling policy in terms of minimizing the average completion time of tasks.
	\item \textbf{Exploration phase:} An idle server is scheduled to one of its non-empty sub-queues uniformly at random, and stays idle if all its sub-queues are empty.
\end{itemize}
Since the arrival rate of any task type is strictly positive, infinitely many of each task type arrives to system, and given the fact that the probability of exploration in both routing and scheduling policies decays such that $\sum_{t = 0}^\infty p(t) = \infty$, using the second Borel-Cantelli lemma (zero-one law), it is obvious that $n$-local tasks to server $m$ are scheduled to this server for infinitely many times for any locality level and any server, so $\widetilde{B}_\mu(t) \hspace{-0.539mm} \rightarrow \hspace{-0.539mm} B_\mu$ as $t \hspace{-0.539mm} \rightarrow \hspace{-0.539mm} \infty$ using the updates in \eqref{estimators}.

\textbf{Remark.} There has been a debate in the queueing community whether the exploration phase in a load balancing algorithm is required to stabilize a queueing system with unknown processing rates or the processing rates are learnt through a natural learning phenomena; and as a result, no exploration is needed.
We provide an example in Figure \ref{counter_example} that shows no exploration can not only increase the mean task completion time, but it can also make the system unstable when the arrival rates are inside the capacity region of the queueing system.
Consider a queueing system as depicted on the left-hand-side of Figure \ref{counter_example}, where the processing times of any tasks on any servers are deterministic with the given rates and the arrival process of tasks is deterministic as well with the rates shown in the figure.
It is obvious that the optimum load balancing is to process task type 1 on server 1 and task type 2 on server 2.
However, if the processing rates are initialized as in the middle queueing system of Figure \ref{counter_example}, for any $\lambda_1 \leq 0.5$ and $\lambda_2 \leq 0.5$, task type 1 is processed by server 2 and task type 2 is processed by server 1 under the GB-PANDAS and MaxWeight algorithms, resulting in a mean task completion time that is two times the optimum value.
On the other hand, if the processing rates are initialized as in the right-hand-side queueing system of Figure \ref{counter_example}, for any $0.5 < \lambda_1 \leq 1$ and $0.5 < \lambda_2 \leq 1$, the system is unstable under the GB-PANDAS and MaxWeight algorithms, while such processing rates are inside the capacity region of the queueing system.
As a result, exploration is required in the load balancing algorithm in general for a queueing system with unknown processing rates. Using the intuition of the given example, it is a promising future work to find conditions for which exploration is not required for the purpose of delay optimality and/or stability.

\subsection{Queue Dynamics under the Blind GB-PANDAS Algorithm}
Denote the queue length vector at time slot $t$ by $\mathbold{Q}(t) = \big ( Q_1^1(t),$ $Q_1^2(t), \cdots, Q_1^{N^1}(t), \cdots, Q_M^{N^M}(t) \big )$.
Let the number of incoming tasks of type $i$ that are routed to their corresponding sub-queue of server $m$ at the beginning of time slot $t$ be denoted as $A_{i, m}(t)$. 
Then, by denoting the number of incoming $n$-local tasks to server $m$ that are routed to $Q_m^n$ at the beginning of time slot $t$ by $A_m^n(t)$, we have:
\begin{equation}
A_m^n(t) = \sum_{i \in \mathcal{L}_m^n} A_{i, m}(t), \ \forall m \in \mathcal{M}, \ 1 \leq n \leq N^m.
\label{sub-queue-arrival}
\end{equation}

Denote the set of working status of servers by vector $\mathbold{f}(t) =$ $\big ( f_1(t),$ $f_2(t),$ $\cdots,$ $f_M(t) \big )$, where
\[
f_m(t) \hspace{-1mm} = \hspace{-1mm} \begin{cases}
			   -1, \ \text{if server $m$ is idle,} \\
                1, \ \text{if server $m$ processes a 1-local task from $Q_m^1$,} \\
			    2, \ \text{if server $m$ processes a 2-local task from $Q_m^2$,} \\
			    \vdots \\
			    N^m, \ \text{if server $m$ processes an $N^m$-local task} \\
			     \text{ \ \ \ \ \ \ from $Q_m^{N^m}$.}
         \end{cases}
\]
If server $m$ finishes processing a task at the end of time slot $t - 1$, i.e. $f_m(t^-) = -1$, a scheduling decision is taken for time $t$ based on $\mathbold{Q}(t)$ and $\mathbold{f}(t)$.
Denote the scheduling decision for server $m$ at time slot $t$ by $\eta_m(t)$ that is defined as follows. For all busy servers, $\eta_m(t) = f_m(t)$, and when $f_m(t^-) = -1$, i.e. server $m$ is idle, $\eta_m(t)$ is determined by the scheduler according to the Blind GB-PANDAS algorithm.
Let $\mathbold{\eta}(t) = \big ( \eta_1(t),$ $\eta_2(t), \cdots, \eta_M(t) \big )$.

Let $S_m^n(t)$ denote the $n$-local service provided by server $m$, where such a service has the rate of $\alpha_m^n$ if $\eta_m(t) = n$ for $1 \leq n \leq N^m$, and the rate is zero otherwise.
Then, the queue dynamics for any $m \in \mathcal{M}$ is as follows:
\begin{equation}
\begin{aligned}
\label{queueevolution}
& Q_m^n(t + 1) = Q_m^n(t) + A_m^n(t) - S_m^n(t), \text{ for } \ 1 \leq \hspace{-0.03cm} n \hspace{-0.05cm} \leq N^m \hspace{-0.05cm} - \hspace{-0.05cm} 1, \\
& Q_m^{N^m}(t + 1) = Q_m^{N^m}(t) + A_m^{N^m}(t) - S_m^{N^m}(t) + U_m(t),
\end{aligned}
\end{equation}
where $U_m(t) = \max \big \{ 0, S_m^{N^m}(t) - A_m^{N^m}(t) - Q_m^{N^m}(t) \big \}$ is the unused service offered by server $m$ at time slot $t$.

Note that $\big \{ \mathbold{Q}(t),$ $t \geq 0 \big \}$ does not necessarily form a Markov chain, i.e. $\mathbold{Q}(t + 1) | \mathbold{Q}(t) \not\perp \mathbold{Q}(t - 1)$, since nothing can be said about locality of an in-service task at a server by just knowing the queue lengths. Even $\big \{ \left ( \mathbold{Q}(t), \mathbold{\eta}(t) \right ),$ $t \geq 0 \big \}$ is not a Markov chain since the service time distributions do not necessarily have the memory-less property.
In order to use Foster-Lyapunov theorem for proving the positive recurrence of a Markov chain, we need to consider another measurement of the status of servers as follows.
\begin{itemize}[leftmargin=*]
	\item Let $\Psi_m(t)$ denote the number of time slots at the beginning of time slot $t$ that server $m$ has been allocated on the current in-service task on server $m$.
	This parameter is set to zero when server $m$ finishes processing a task. Let $\mathbold{\Psi}(t) = \big ( \Psi_1(t), \Psi_2(t), \cdots, \Psi_M(t) \big )$.
\end{itemize}

\begin{lemma}
\label{markov_chain}
$\big \{ \mathbold{Z}(t) = \big (\mathbold{Q}(t), \mathbold{\eta}(t), \mathbold{\Psi}(t) \big ), t \geq 0 \big \}$ forms an irreducible and aperiodic Markov chain. The state space of this Markov chain is $\mathcal{S} = \big ( \prod_{m \in \mathcal{M}} \mathbb{N}^{N^m} \big ) \times \big ( \prod_{m \in \mathcal{M}} \{ 1, 2,$ $\cdots,$ $N^m \} \big ) \times \mathbb{N}^{M}$.
\end{lemma}


\section{Throughput Optimality of the Blind GB-PANDAS Algorithm}
\label{ThroughputOptimalitySection}
Section \ref{section1} provides preliminaries on the workload dynamic of servers, the ideal workload on servers, some lemmas, and an extended version of the Foster-Lyapunov.
The throughput-optimality theorem of the Blind GB-PANDAS algorithm and its proof are presented in Section \ref{section2}, where the proof is followed by using Lemmas \ref{newlemma2}, \ref{lemma123}, \ref{lemma1234}, \ref{lemma12345}, and \ref{lemma123456}.

\subsection{Preliminary Materials and Lemmas}

The workload on server $m$ evolves as follows:
\[
\begin{aligned}
& W_m (t + 1 ) = \frac{Q_m^1(t+1)}{\alpha_m^1} + \frac{Q_m^2(t+1)}{\alpha_m^2} + \cdots + \frac{Q_m^{N^m}(t+1)}{\alpha_m^{N^m}}  \\
& \hspace{0.25cm} \overset{(a)}{=} \frac{Q_m^1(t) + A_m^1(t) - S_m^1(t)}{\alpha_m^1} + \frac{Q_m^2(t) + A_m^2(t) - S_m^2(t)}{\alpha_m^2} + \\
& \hspace{2.25cm} \cdots + \frac{Q_m^{N^m}(t) + A_m^{N^m}(t) - S_m^{N^m}(t) + U_m(t)}{\alpha_m^{N^m}} \\
& \hspace{0.25cm} = W_m(t) + \bigg (\frac{A_m^1(t)}{\alpha_m^1} + \frac{A_m^2(t)}{\alpha_m^2} + \cdots + \frac{A_m^{N^m}(t)}{\alpha_m^{N^m}} \bigg ) \\
&  \hspace{1.7cm} - \bigg (\frac{S_m^1(t)}{\alpha_m^1} + \frac{S_m^2(t)}{\alpha_m^2} + \cdots + \frac{S_m^{N^m}(t)}{\alpha_m^{N^m}} \bigg ) + \frac{U_m(t)}{\alpha_m^{N^m}} \\
& \hspace{0.25cm} \overset{(b)}{=} W_m(t) + A_m(t) - S_m(t) + \widetilde{U}_m(t),
\end{aligned}
\]
where $(a)$ is true by using the queue dynamics in \eqref{queueevolution} and $(b)$ follows from defining the pseudo task arrival, service, and unused services of server $m$ as
\begin{equation}
\begin{aligned}
A_m(t) & = \frac{A_m^1(t)}{\alpha_m^1} + \frac{A_m^2(t)}{\alpha_m^2} + \cdots + \frac{A_m^{N^m}(t)}{\alpha_m^{N^m}}, \ \forall m \in \mathcal{M}, \\
S_m(t) & = \frac{S_m^1(t)}{\alpha_m^1} + \frac{S_m^2(t)}{\alpha_m^2} + \cdots + \frac{S_m^{N^m}(t)}{\alpha_m^{N^m}}, \ \forall m \in \mathcal{M}, \\
\widetilde{U}_m(t) & = \frac{U_m(t)}{\alpha_m^{N^m}}, \ \forall m \in \mathcal{M}.
\label{pseudoparameters}
\end{aligned}
\end{equation}
By defining the pseudo task arrival, service, and unused service processes as $\mathbold{A}(t) = \big (A_1(t), A_2(t), \cdots, A_M(t) \big )$, $\mathbold{S}(t) = \big (S_1(t), S_2(t),$ $\cdots, S_M(t) \big )$, and $\widetilde{\mathbold{U}}(t) =$ $\big (\widetilde{U}_1(t), \widetilde{U}_2(t),$ $\cdots,$ $\widetilde{U}_M(t) \big )$, respectively, the vector of servers' workloads defined by $\mathbold{W} = (W_1, W_2, \cdots, W_M)$ evolves as
\begin{equation}
\label{evolW}
\mathbold{W}(t + 1) = \mathbold{W}(t) + \mathbold{A}(t) - \mathbold{S}(t) + \widetilde{\mathbold{U}}(t).
\end{equation}

\begin{lemma}
\label{newlemma2}
For any arrival rate vector inside the capacity region, $\mathbold{\lambda} \in \Lambda$, there exists a load decomposition $\{\lambda_{i, m}\}$ and $\delta > 0$ such that
\begin{equation}
\label{optimum_load_decomposition}
\sum_{i \in \mathcal{L}} \frac{\lambda_{i, m}}{\mu_{i, m}} < \frac{1}{1 + \delta}, \ \forall m \in \mathcal{M}.
\end{equation}
The fluid model planning algorithm solves a linear programming to find the load decomposition $\{\lambda_{i, m}\}$ that is used in its load balancing on the $M$ servers. In  other words, this load decomposition is a possibility of task assignment on servers to stabilize the system.
\end{lemma}
Lemma \ref{newlemma2} is used in the proof of Lemma
 \ref{lemma12345}.

\begin{definition}
The ideal workload on server $m$ corresponding to the load decomposition $\{\lambda_{i, m}\}$ of Lemma \ref{newlemma2} is defined as
\begin{equation}
\label{workloadm}
w_m = \sum_{i \in \mathcal{L}} \frac{\lambda_{i, m}}{\mu_{i, m}}, \ \forall m \in \mathcal{M}.
\end{equation}
Let $\mathbold{ w} = (w_1, w_2, \cdots, w_M )$. The vector of servers' ideal workload is used as an intermediary term in Lemmas \ref{lemma1234} and \ref{lemma12345} which are later used for throughput optimality proof of the Blind GB-PANDAS algorithm.
\end{definition}

\begin{lemma}
\label{lemma123}
\begin{equation*}
\langle \mathbold{W}(t) , \widetilde{\mathbold{U}}(t) \rangle = 0, \ \forall t.
\end{equation*}
\end{lemma}

The following lemma states that the sum over a time period of the inner product of the workload and the pseudo arrival rate is dominated on an expectation sense by the inner product of the workload and the ideal workload plus constants depending on the initial state of the system.
\begin{lemma}
\label{lemma1234}
Under the exploration-exploitation routing policy of the Blind GB-PANDAS algorithm, for any arrival rate vector inside the capacity region, $\mathbold{\lambda} \in \Lambda$, and the corresponding ideal workload vector $\mathbold{w}$ defined in  \eqref{workloadm}, and for any arbitrary small $\theta_0 > 0$, there exists $T_0 > t_0$ such that for any $t_0 \geq 0$ and $T > T_0$:
\[
\begin{aligned}
& \mathbb{E} \Big [ \sum_{t = T_0}^{t_0 + T - 1} \Big ( \langle \mathbold{W}(t), \mathbold{A}(t) \rangle - \langle \mathbold{W}(t), \mathbold{w} \rangle \Big ) \Big | \mathbold{Z}(t_0) \Big ] \\
& \ \ \ \ \ \ \ \ \ \ \ \ \ \ \ \ \ \ \ \ \ \ \ \ \ \ \ \ \ \ \ \ \ \ \ \ \ \ \ \ \ \ \ \ \ \ \ \ \ \ \ \ \ \ \leq \theta_0 T \| \mathbold{Q}(t_0) \|_1 + c_0,
\end{aligned}
\]
where the constants $\theta_0, c_0 > 0$ are independent of $\mathbold{Z}(t_0)$.
\end{lemma}

We emphasize that $\theta_0$ in Lemma \ref{lemma1234} can be made arbitrarily small, as can be seen in the proof, which is used in the throughput optimality proof of Blind GB-PANDAS, Theorem  \ref{theorem_th}. Throughout this paper, $\|.\|$ and $\|.\|_1$ are the $L^2$-norm and $L^1$-norm, respectively.

The following lemma is the counterpart of Lemma \ref{lemma1234} for the pseudo service process.
\begin{lemma}
\label{lemma12345}
Under the exploration-exploitation scheduling policy of the Blind GB-PANDAS algorithm, for any arrival rate vector inside the capacity region, $\mathbold{\lambda} \in \Lambda$, and the corresponding ideal workload vector $\mathbold{w}$ in \eqref{workloadm}, there exists $T_1 > 0$ such that for any $T > T_1$, we have:
\begin{equation}
\begin{aligned}
& \mathbb{E} \left [ \sum_{t = t_0}^{t_0 + T - 1} \Big ( \langle \mathbold{W}(t), \mathbold{w} \rangle - \langle \mathbold{W}(t), \mathbold{S}(t) \rangle \Big ) \Big | \mathbold{Z}(t_0) \right ] \\
& \ \ \ \ \ \ \ \ \ \ \ \ \ \ \ \ \ \ \ \ \ \ \ \ \ \ \ \leq - \theta_1 T \| {\mathbold{Q}}(t_0) \|_1 + c_1, \ \forall t_0 \geq 0,
\label{IIII}
\end{aligned}
\end{equation}
where the constants $\theta_1, c_1 > 0$ are independent of $\mathbold{Z}(t_0)$.
\end{lemma}

\begin{lemma}
\label{lemma123456}
Under the exploration-exploitation load balancing of the Blind GB-PANDAS algorithm, for any arrival rate vector inside the capacity region, $\mathbold{\lambda} \in \Lambda$, and for any $\theta_2 > 0$, there exists $T_2 > 0$ such that for any $T > T_2$ and for any $t_0 \geq 0$, we have:
\[
\hspace{-0.12cm} \mathbb{E} \Big [ \|\mathbold{\Psi}(t_0 + T)\|_1 - \|\mathbold{\Psi}(t_0)\|_1 \Big | \mathbold{Z}(t_0) \Big ] \leq -\theta_2 \|\mathbold{\Psi}(t_0)\|_1 + M T.
\]
\end{lemma}

Theorem 3.3.8 in \cite{srikant2013communication}, an extended version of the Foster-Lyapunov theorem: Consider an irreducible Markov chain $\{ \mathbold{Z}(t) \}$, where $t \in \mathbb{N}$, with a state space $\mathcal{S}$. If there exists a function $V: \mathcal{S} \rightarrow \mathcal{R}^+$, a positive integer $T \geq 1$, and a finite set $\mathcal{P} \subseteq \mathcal{S}$ satisfying the following condition:
\begin{equation}
\begin{aligned}
& \mathbb{E} \big [ V(Z(t_0 + T)) - V(Z(t_0)) \big | \mathbold{Z}(t_0) = z  \big ] \\
& \hspace{4.129cm} \leq - \theta \mathbb{I}_{\{ z \in \mathcal{P}^c \}} + C \mathbb{I}_{\{ z \in \mathcal{P} \}},
\label{extendedLyapunov}
\end{aligned}
\end{equation}
for some $\theta > 0$ and $C < \infty$, then the irreducible Markov chain $\{ \mathbold{Z}(t) \}$ is positive recurrent.

\label{section1}

\subsection{Throughput Optimality Theorem and Proof}
\label{section2}
\begin{theorem}
\label{theorem_th}
The Blind GB-PANDAS algorithm is throughput-optimal for a system with affinity setup discussed in Section \ref{SystemModelSection}, with general service time distributions with finite means and supports, without prior knowledge on the service rate matrix $B_\mu$ and the arrival rate vector $\mathbold{\lambda}$.
\end{theorem}

\textbf{Proof:} We use the Foster-Lyapunov theorem for proving that the irreducible and aperiodic Markov chain $\big \{ \mathbold{Z}(t) = \big ( \mathbold{Q}(t), \mathbold{\eta}(t), \mathbold{\Psi}(t) \big ),$ $t \geq 0 \big \}$ (Lemma \ref{markov_chain}) is positive recurrent under the Blind GB-PANDAS algorithm, as far as the arrival rate vector is inside the capacity region, $\mathbold{\lambda} \in \Lambda$. This means that as time goes to infinity, the distribution of $\mathbold{Z}(t)$ converges to its stationary distribution, which implies that the system is stable and Blind GB-PANDAS is throughput-optimal. To this end, we choose the following Lyapunov function $V:\mathcal{S} \rightarrow \mathcal{R}^+$ and use Lemmas \ref{newlemma2}, \ref{lemma123}, \ref{lemma1234}, \ref{lemma12345}, and \ref{lemma123456} to derive its drift afterward:
\begin{equation}
\label{Lyapunov_function_1}
V(\mathbold{Z}(t)) = \| \mathbold{W}(t) \|^2 + \|\mathbold{\Psi}(t)\|_1.
\end{equation}
By choosing $\theta_0$ in Lemma \ref{lemma1234} less than $\theta_1$ in Lemma \ref{lemma12345}, $\theta_0 < \theta_1$, we get $T_0$ from Lemma \ref{lemma1234}, which is used in the drift of the Lyapunov function in Lemma \ref{lemma_drift}.
\begin{lemma}
\label{lemma_drift}
For any $t_0 \leq T_0 < T$, specifically $T_0$ from Lemma \ref{lemma1234} that is dictated by choosing $\theta_0 < \theta_1$, we have the following for the drift of the Lyapunov function in \eqref{Lyapunov_function_1}, where $T_0$ is used in the first summation after the inequality:
\begin{equation}
\begin{aligned}
& \mathbb{E} \Big [ V(\mathbold{Z}(t_0 + T)) - V(\mathbold{Z}(t_0)) \Big | \mathbold{Z}(t_0) \Big ] \\
\leq & 2 \mathbb{E} \left [ \sum_{t = T_0}^{t_0 + T - 1} \Big ( \langle \mathbold{W}(t), \mathbold{A}(t) \rangle - \langle \mathbold{W}(t), \mathbold{w} \rangle \Big ) \Big | \mathbold{Z}(t_0) \right ] \\
& \hspace{-0.35cm} + 2 \mathbb{E} \left [ \sum_{t = t_0}^{t_0 + T - 1} \Big ( \langle \mathbold{W}(t), \mathbold{w} \rangle - \langle \mathbold{W}(t), \mathbold{S}(t) \rangle \Big ) \Big | \mathbold{Z}(t_0) \right ] \\
& \hspace{-0.35cm} + \mathbb{E} \Big [ \| \mathbold{\Psi}(t_0 + T) \|_1 - \| \mathbold{\Psi}(t) \|_1 \Big | \mathbold{Z}(t_0) \Big ] + c_2 \| \mathbold{Q}(t_0) \|_1 + c_3.
\label{driftlyapunov_1}
\end{aligned}
\end{equation}
\end{lemma}
By choosing $T > \max\{ T_0, T_1, T_2, \frac{\theta_2 + c_2}{2(\theta_1 - \theta_0)} \}$, where $\theta_2 > 0$ is the one in Lemma \ref{lemma123456}, and substituting the terms on the right-hand side of the Lyapunov function drift \eqref{driftlyapunov_1} in Lemma \ref{lemma_drift} from the corresponding inequalities in Lemmas \ref{lemma1234}, \ref{lemma12345}, and \ref{lemma123456}, we have:
\[
\begin{aligned}
& \mathbb{E} \Big [ V(\mathbold{Z}(t_0 + T)) - V(\mathbold{Z}(t_0)) \Big | \mathbold{Z}(t_0) \Big ] \\
\leq & -\theta_2 \Big ( \|\mathbold{Q}(t_0)\|_1 + \|\mathbold{\Psi}(t_0)\|_1 \Big ) + c, \ \forall t_0,
\end{aligned}
\]
where $c = 2c_0 + 2c_1 + c_3 + MT$.\\
Let $\mathcal{P} = \big \{ \mathbold{Z} = \big (\mathbold{Q}, \mathbold{\eta}, \mathbold{\Psi} \big ) \in \mathcal{S} : \|\mathbold{Q}\|_1 + \|\mathbold{\Psi}\|_1 \leq \frac{\bar{c} + c}{\theta_2} \big \}$ for any positive constant $\bar{c} > 0$, where $\mathcal{P}$ is a finite set of the state space $\mathcal{S}$. By this choice of $\mathcal{P}$ for the Lyapunov function $V(.)$ defined in \eqref{Lyapunov_function_1}, all the conditions of the Foster-Lyapunov theorem are satisfied, which completes the throughput optimality proof for the Blind GB-PANDAS algorithm.

Note that the priority scheduling in the exploitation phase of the Blind GB-PANDAS algorithm is not used for the throughput optimality proof since the expected workload of a server is decreased in the same rate no matter what locality level is receiving service from the server. As long as an idle server gives service to one of the tasks in its sub-queues continuously, the system is stable. Given the routing policy, the priority scheduling is used in the exploitation phase to minimize the mean task completion time. 

\section{Simulation Results}
\label{SimulationResultsSection}

\begin{figure}[t]
\centering
\includegraphics[scale=0.6]{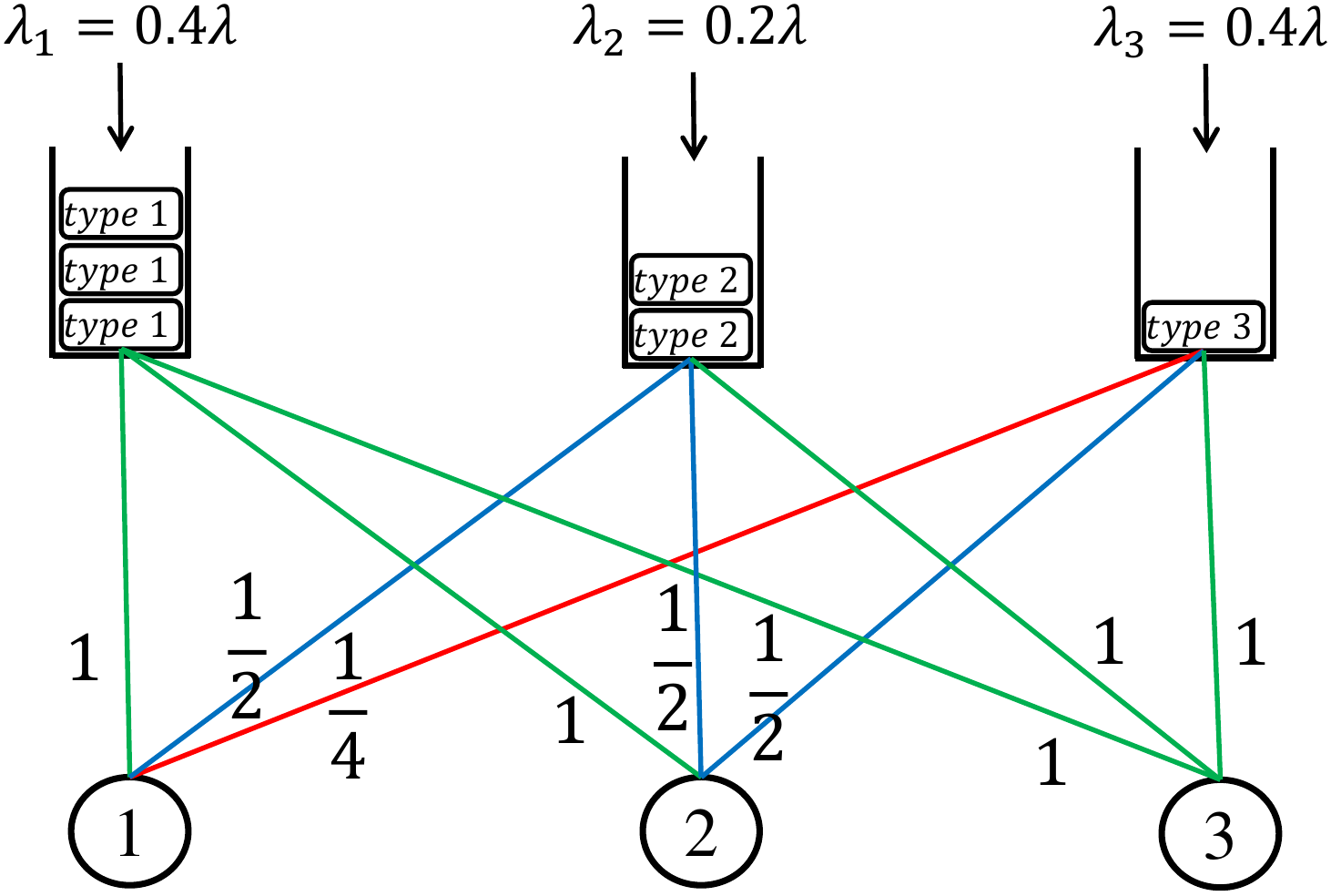}
\caption{The affinity structure used for simulation with three types of tasks and three multi-skilled servers.}
\label{simulated_system}
\end{figure}

In this section, the simulated performance of the Blind GB-PANDAS algorithm is compared with FCFS, Max-Weight, and c-$\mu$-rule algorithms.
FCFS does not use system parameters for load balancing, but Max-Weight and c-$\mu$-rule use the same exploration-exploitation approach as Blind GB-PANDAS. Convex cost functions $C_i(Q_i) = Q_i^{1.01}$ for $i \in \{1, 2, 3\}$ are used for the c-$\mu$-rule algorithm. Since the objective is to minimize the mean task completion time, the convexity of the cost functions are chosen in a way to be close to a line for small values of $Q_i$. Three types of tasks and a computing cluster of three servers are considered with processing rates depicted in Figure \ref{simulated_system}, which are not known from the perspective of the load balancing algorithms.
The task arrivals are Poisson processes with the unknown rates determined in Figure \ref{simulated_system} and the processing times are log-normal that are heavy-tailed and do not have the memory-less property.
Note that this affinity structure does not have the rack structure mentioned in \cite{xie2016scheduling} since from the processing rates of task type $2$ on the three servers, servers $1$ and $2$ are in the same rack as server $3$, but from the processing rates of task type $3$ on the three servers, the second server is in the same rack as the third server, but not the first server.
Hence, this affinity setup is more complicated than the one with a rack structure.

\begin{figure}[t]
\centering
\includegraphics[scale=0.1578]{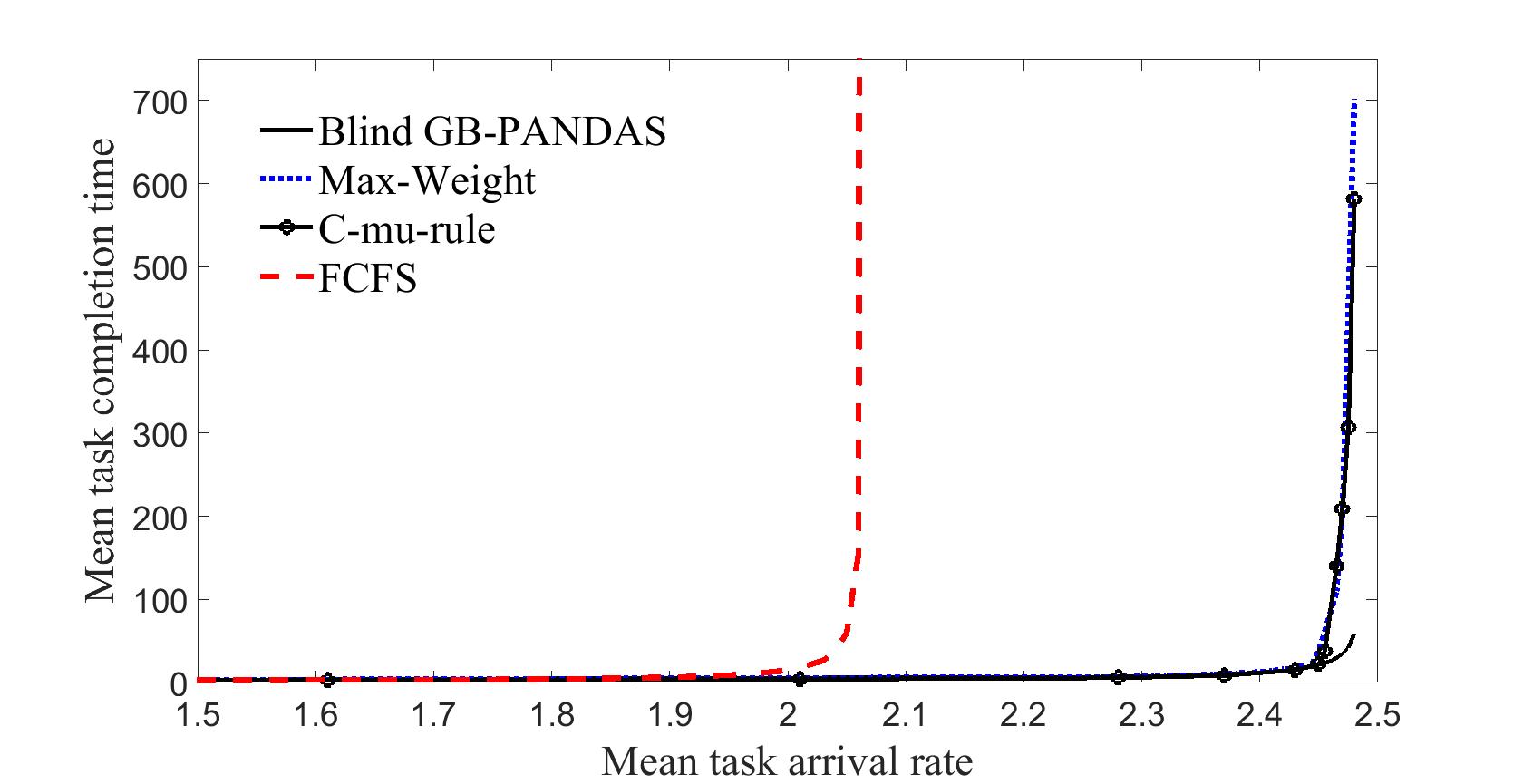}
\caption{Capacity region comparison of the Blind GB-PANDAS, Max-Weight, c-$\mu$-rule, and FCFS algorithms.}
\label{comparison0}
\end{figure}
\begin{figure}[t]
\centering
\includegraphics[scale=0.1578]{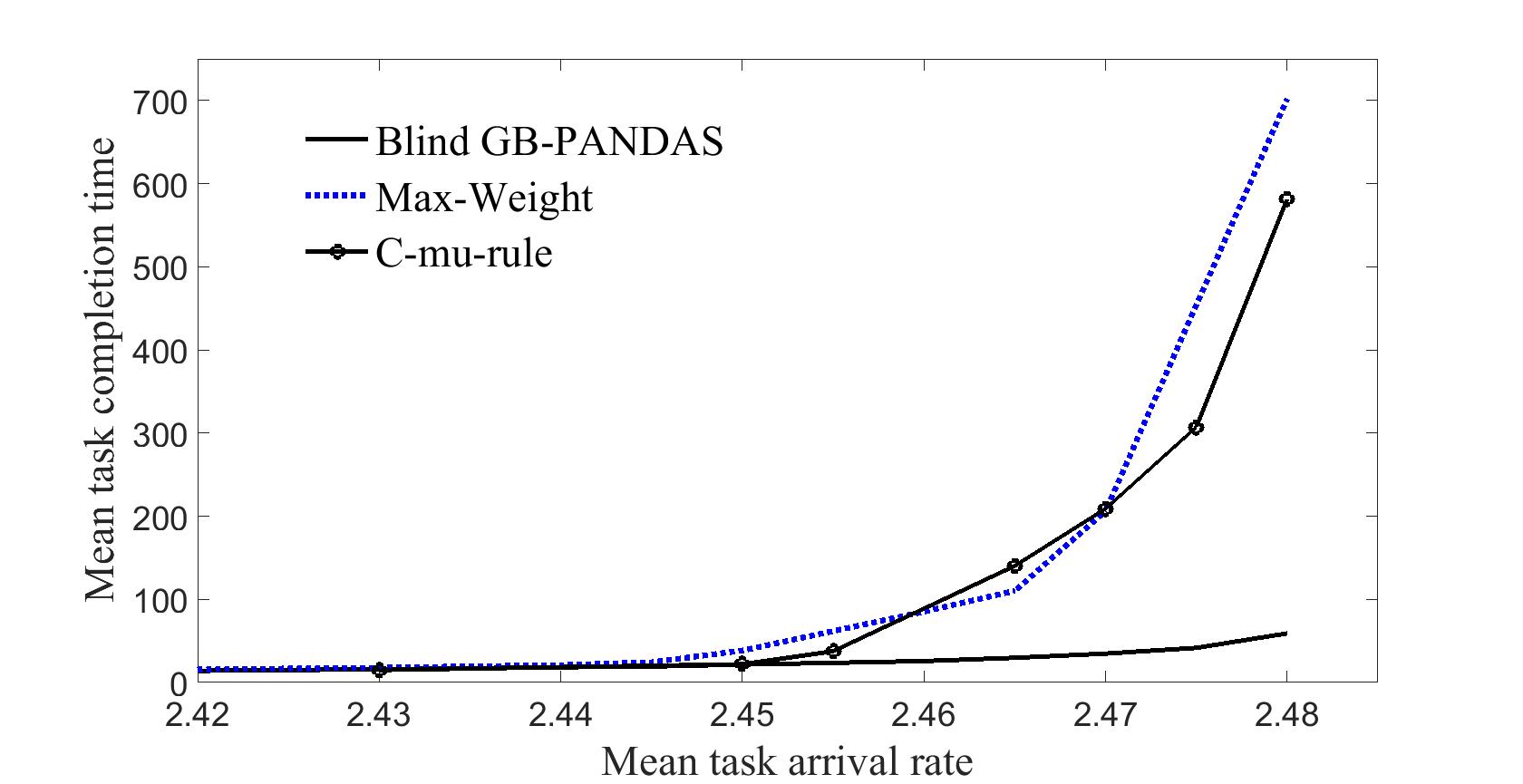}
\caption{Heavy-traffic performance comparison.}
\label{comparison}
\end{figure}

Inspired by the fluid model planning algorithm, the following linear programming optimization should be solved to find the capacity region of the simulated system.
\[
\begin{aligned}
& \underset{\lambda_{i, m}}{\text{maximize}} \ \ \lambda = \sum_{i = 1}^3 \sum_{m = 1}^3 \lambda_{i, m} \\
& \text{subject to:} \\
& \ \ \ \ \ \ \lambda_{1, 1} + 2 \lambda_{2, 1} + 4 \lambda_{3, 1} < 1, \ \ \ \ \ \ \lambda_{1, 1} + \lambda_{1, 2} + \lambda_{1, 3} = 0.4 \lambda, \\
& \ \ \ \ \ \ \lambda_{1, 2} + 2 \lambda_{2, 2} + 2 \lambda_{3, 2} < 1, \ \ \ \ \ \ \lambda_{2, 1} + \lambda_{2, 2} + \lambda_{2, 3} = 0.2 \lambda, \\
& \ \ \ \ \ \ \lambda_{1, 3} + \lambda_{2, 3} + \lambda_{3, 3} < 1, \ \ \ \ \ \ \ \ \ \ \lambda_{3, 1} + \lambda_{3, 2} + \lambda_{3, 3} = 0.4 \lambda, \\
& \ \ \ \ \ \ \ \ \ \ \ \ \ \ \ \ \ \ \ \ \ \ \ \ \ \ \ \lambda_{i, m} \geq 0, \ \forall i, m \in \{1, 2, 3\}. \\
\end{aligned}
\]
The capacity region in terms of $\lambda$ is found to be $\lambda \in [0, 2.5)$. Figure \ref{comparison0} compares the throughput performance of the four algorithms, where the mean task completion time versus the total task arrival rate, $\lambda = \sum_{i = 1}^3 \lambda_i$, is plotted. The Blind GB-PANDAS, Max-Weight, and c-$\mu$-rule algorithms are throughput-optimal by stabilizing the system for $\lambda < 2.5$. Taking a closer look at the performance of these algorithms at high loads, Blind GB-PANDAS has a much lower mean task completion time compared to Max-Weight and c-$\mu$-rule algorithms as depicted in Figure \ref{comparison}.

\section{Related Work}
\label{RelatedWorkSection}
In addition to the fluid model planning, Max-Weight, and c-$\mu$-rule algorithms for the affinity scheduling problem that are discussed in Section \ref{introduction} and Section \ref{MWCMU}, there is a huge body of work on heuristic algorithms that are used for scheduling for data centers with multiple levels of data locality, e.g. \cite{white2010hadoop
, isard2009quincy, zaharia2010delay
}, and look at the references in \cite{yekkehkhany2017gb}.
 Although some of these heuristic algorithms are being used in real applications, simple facts about their optimality are not investigated.
Recent works including the priority algorithm \cite{xie2015priority}, Join-the-Shortest-Queue-Max-Weight (JSQ-MW) \cite{wang2016maptask} and Weighted-Workload algorithm \cite{xie2016scheduling} study the capacity region and throughput optimality for a system with two and three levels of data locality.
Some robust policies are studied in \cite{lu1991distributed, dai1996stability, baharian2011stability, dimakis2006sufficient, pedarsani2016stability}.

A related direction of work on scheduling for data centers with multi-level data locality, which is a direct application of affinity scheduling, is to efficiently do data replication on servers in MapReduce framework to increase availability. Increasing the availability is translated to increasing service rates in the context of this article which enlarges the capacity region and reduces the mean task completion time. For more information on data replication algorithms refer to Google File System \cite{sanjay2003google}, Hadoop Distributed File System \cite{white2010hadoop}, Scarlett \cite{ananthanarayanan2011scarlett}, and Dare \cite{abad2011dare}. Data replication algorithms are complementary and orthogonal to throughput and delay optimality that is studied in this article.

Fairness is an issue in most scheduling problems which conflicts with delay optimality. A delay optimal load balancing algorithm can cooperate with fair scheduling strategies though by compromising on delay optimality to partly achieve fairness. Refer to Delay Scheduling \cite{zaharia2010delay}, Quibcy \cite{isard2009quincy}, and the references therein for more details.

\section{Conclusion and Future Work}
\label{ConclutionSection}
The Blind GB-PANDAS algorithm is proposed for the affinity load balancing problem where no knowledge of the task arrival rates and the service rate matrix is available. An exploration-exploitation approach is proposed for load balancing which consists of exploration and exploitation phases. The system is proven to be stable under Blind GB-PANDAS and is shown empirically through simulations to have a better delay performance 
than Max-Weight, c-$\mu$-rule, and FCFS algorithms. Investigating the subspace of the capacity region in which GB-PANDAS is delay optimal is a promising direction for future work. Note that both GB-PANDAS and Max-Weight algorithms have high routing and scheduling computation complexity which can be alleviated using power-of-d-choices \cite{mukherjee2016universality} or join-idle-queue \cite{lu2011join} algorithms which are interesting directions to study as well.
Another interesting future work is to consider a case where there are precedence relations between several tasks of a job, i.e. a departing task may join another queue.


\bibliographystyle{IEEEtran}
\bibliography{sigproc}

\begin{thebibliography}{10}
\providecommand{\url}[1]{#1}
\csname url@samestyle\endcsname
\providecommand{\newblock}{\relax}
\providecommand{\bibinfo}[2]{#2}
\providecommand{\BIBentrySTDinterwordspacing}{\spaceskip=0pt\relax}
\providecommand{\BIBentryALTinterwordstretchfactor}{4}
\providecommand{\BIBentryALTinterwordspacing}{\spaceskip=\fontdimen2\font plus
\BIBentryALTinterwordstretchfactor\fontdimen3\font minus
  \fontdimen4\font\relax}
\providecommand{\BIBforeignlanguage}[2]{{%
\expandafter\ifx\csname l@#1\endcsname\relax
\typeout{** WARNING: IEEEtran.bst: No hyphenation pattern has been}%
\typeout{** loaded for the language `#1'. Using the pattern for}%
\typeout{** the default language instead.}%
\else
\language=\csname l@#1\endcsname
\fi
#2}}
\providecommand{\BIBdecl}{\relax}
\BIBdecl

\bibitem{padua2011encyclopedia}
D.~Padua, \emph{Encyclopedia of parallel computing}.\hskip 1em plus 0.5em minus
  0.4em\relax Springer Science \& Business Media, 2011.

\bibitem{harrison1998heavy}
J.~M. Harrison, ``Heavy traffic analysis of a system with parallel servers:
  Asymptotic optimality of discrete-review policies,'' \emph{Annals of Applied
  Probability}, pp. 822--848, 1998.

\bibitem{harrison1999heavy}
J.~M. Harrison and M.~J. L{\'o}pez, ``Heavy traffic resource pooling in
  parallel-server systems,'' \emph{Queueing Systems}, vol.~33, no.~4, pp.
  339--368, 1999.

\bibitem{bell2001dynamic}
S.~L. Bell, R.~J. Williams \emph{et~al.}, ``Dynamic scheduling of a system with
  two parallel servers in heavy traffic with resource pooling: asymptotic
  optimality of a threshold policy,'' \emph{The Annals of Applied Probability},
  vol.~11, no.~3, pp. 608--649, 2001.

\bibitem{bell2005dynamic}
S.~Bell, R.~Williams \emph{et~al.}, ``Dynamic scheduling of a parallel server
  system in heavy traffic with complete resource pooling: Asymptotic optimality
  of a threshold policy,'' \emph{Electronic Journal of Probability}, vol.~10,
  pp. 1044--1115, 2005.

\bibitem{xie2016scheduling}
Q.~Xie, A.~Yekkehkhany, and Y.~Lu, ``Scheduling with multi-level data locality:
  Throughput and heavy-traffic optimality,'' in \emph{Computer Communications,
  IEEE INFOCOM 2016-The 35th Annual IEEE International Conference on}.\hskip
  1em plus 0.5em minus 0.4em\relax IEEE, 2016, pp. 1--9.

\bibitem{yekkehkhany2017near}
A.~Yekkehkhany, ``Near data scheduling for data centers with multi levels of
  data locality,'' \emph{(Dissertation, University of Illinois at
  Urbana-Champaign)}.

\bibitem{stolyar2004maxweight}
A.~L. Stolyar, ``Maxweight scheduling in a generalized switch: State space
  collapse and workload minimization in heavy traffic,'' \emph{Annals of
  Applied Probability}, pp. 1--53, 2004.

\bibitem{mandelbaum2004scheduling}
A.~Mandelbaum and A.~L. Stolyar, ``Scheduling flexible servers with convex
  delay costs: Heavy-traffic optimality of the generalized c$\mu$-rule,''
  \emph{Operations Research}, vol.~52, no.~6, pp. 836--855, 2004.

\bibitem{wang2016maptask}
W.~Wang, K.~Zhu, L.~Ying, J.~Tan, and L.~Zhang, ``Maptask scheduling in
  mapreduce with data locality: Throughput and heavy-traffic optimality,''
  \emph{IEEE/ACM Transactions on Networking}, vol.~24, no.~1, pp. 190--203,
  2016.

\bibitem{yekkehkhany2017gb}
A.~Yekkehkhany, A.~Hojjati, and M.~H. Hajiesmaili, ``Gb-pandas:: Throughput and
  heavy-traffic optimality analysis for affinity scheduling,'' \emph{ACM
  SIGMETRICS Performance Evaluation Review}, vol.~45, no.~2, pp. 2--14, 2018.

\bibitem{pedarsani2017robust}
R.~Pedarsani, J.~Walrand, and Y.~Zhong, ``Robust scheduling for flexible
  processing networks,'' \emph{Advances in Applied Probability}, vol.~49,
  no.~2, pp. 603--628, 2017.

\bibitem{srikant2013communication}
R.~Srikant and L.~Ying, \emph{Communication networks: an optimization, control,
  and stochastic networks perspective}.\hskip 1em plus 0.5em minus 0.4em\relax
  Cambridge University Press, 2013.

\bibitem{white2010hadoop}
T.~White, ``Hadoop: The definitive guide, yahoo,'' 2010.

\bibitem{isard2009quincy}
M.~Isard, V.~Prabhakaran, J.~Currey, U.~Wieder, K.~Talwar, and A.~Goldberg,
  ``Quincy: Fair scheduling for distributed computing clusters,'' in
  \emph{Proceedings of the ACM SIGOPS 22nd Symposium on Operating Systems
  Principles}.\hskip 1em plus 0.5em minus 0.4em\relax ACM, 2009, pp. 261--276.

\bibitem{zaharia2010delay}
M.~Zaharia, D.~Borthakur, J.~Sen~Sarma, K.~Elmeleegy, S.~Shenker, and
  I.~Stoica, ``Delay scheduling: A simple technique for achieving locality and
  fairness in cluster scheduling,'' in \emph{Proceedings of the 5th European
  Conference on Computer Systems}.\hskip 1em plus 0.5em minus 0.4em\relax ACM,
  2010, pp. 265--278.

\bibitem{xie2015priority}
Q.~Xie and Y.~Lu, ``Priority algorithm for near-data scheduling: Throughput and
  heavy-traffic optimality,'' in \emph{Computer Communications (INFOCOM), 2015
  IEEE Conference on}.\hskip 1em plus 0.5em minus 0.4em\relax IEEE, 2015, pp.
  963--972.

\bibitem{lu1991distributed}
S.~H. Lu and P.~Kumar, ``Distributed scheduling based on due dates and buffer
  priorities,'' \emph{IEEE Transactions on Automatic Control}, vol.~36, no.~12,
  pp. 1406--1416, 1991.

\bibitem{dai1996stability}
J.~G. Dai and G.~Weiss, ``Stability and instability of fluid models for
  reentrant lines,'' \emph{Mathematics of Operations Research}, vol.~21, no.~1,
  pp. 115--134, 1996.

\bibitem{baharian2011stability}
G.~Baharian and T.~Tezcan, ``Stability analysis of parallel server systems
  under longest queue first,'' \emph{Mathematical Methods of Operations
  Research}, vol.~74, no.~2, p. 257, 2011.

\bibitem{dimakis2006sufficient}
A.~Dimakis and J.~Walrand, ``Sufficient conditions for stability of
  longest-queue-first scheduling: Second-order properties using fluid limits,''
  \emph{Advances in Applied probability}, vol.~38, no.~2, pp. 505--521, 2006.

\bibitem{pedarsani2016stability}
R.~Pedarsani and J.~Walrand, ``Stability of multiclass queueing networks under
  longest-queue and longest-dominating-queue scheduling,'' \emph{Journal of
  Applied Probability}, vol.~53, no.~2, pp. 421--433, 2016.

\bibitem{sanjay2003google}
G.~Sanjay, G.~Howard, and L.~Shun-Tak, ``The google file system,'' in
  \emph{Proceedings of the 17th ACM Symposium on Operating Systems Principles},
  2003, pp. 29--43.

\bibitem{ananthanarayanan2011scarlett}
G.~Ananthanarayanan, S.~Agarwal, S.~Kandula, A.~Greenberg, I.~Stoica,
  D.~Harlan, and E.~Harris, ``Scarlett: coping with skewed content popularity
  in mapreduce clusters,'' in \emph{Proceedings of the sixth conference on
  Computer systems}.\hskip 1em plus 0.5em minus 0.4em\relax ACM, 2011, pp.
  287--300.

\bibitem{abad2011dare}
C.~L. Abad, Y.~Lu, and R.~H. Campbell, ``Dare: Adaptive data replication for
  efficient cluster scheduling,'' in \emph{Cluster Computing (CLUSTER), 2011
  IEEE International Conference on}.\hskip 1em plus 0.5em minus 0.4em\relax
  Ieee, 2011, pp. 159--168.

\bibitem{mukherjee2016universality}
D.~Mukherjee, S.~C. Borst, J.~S. van Leeuwaarden, and P.~A. Whiting,
  ``Universality of power-of-$ d $ load balancing in many-server systems,''
  \emph{arXiv preprint arXiv:1612.00723}, 2016.

\bibitem{lu2011join}
Y.~Lu, Q.~Xie, G.~Kliot, A.~Geller, J.~R. Larus, and A.~Greenberg,
  ``Join-idle-queue: A novel load balancing algorithm for dynamically scalable
  web services,'' \emph{Performance Evaluation}, vol.~68, no.~11, pp.
  1056--1071, 2011.

\end{thebibliography}


%

\appendix

\subsection{Proof of Lemma \ref{markov_chain}}
Lemma \ref{markov_chain}:
$\big \{ \mathbold{Z}(t) = \big (\mathbold{Q}(t), \mathbold{\eta}(t), \mathbold{\Psi}(t) \big ), t \geq 0 \big \}$ forms an irreducible and aperiodic Markov chain. The state space of this Markov chain is $\mathcal{S} = \big ( \prod_{m \in \mathcal{M}} \mathbb{N}^{N^m} \big ) \times \big ( \prod_{m \in \mathcal{M}} \{ 1, 2,$ $\cdots,$ $N^m \} \big ) \times \mathbb{N}^{M}$.

\textbf{Proof:}
Consider $\mathbold{Z}(0) = \big \{ 0_{(\sum_{m \in \mathcal{M}} N^m ) \times 1}, \prod_{m \in \mathcal{M}} N^m,$ $0_{M \times 1} \big \}$ as the initial state of the Markov chain $\mathbold{Z}(t)$. \\
Irreducible: Since $F_{i, m}$ is increasing for any task-server pair, we can find an integer $\tau > 0$ such that $F_{i, m}(\tau) > 0$ for any $1 \leq i \leq N^m$ and $m \in \mathcal{M}$. Furthermore, probability of zero task arrival is positive in each time slot. Hence, for any state $\mathbold{Z} = (\mathbold{Q}, \mathbold{\eta}, \mathbold{\Psi})$, there is a positive probability that each task receives service in $\tau$ time slots and no new task arrives at the system in $\tau \sum_{m \in \mathcal{M}} \sum_{n = 1}^{N^m} Q_m^n$ time slots. Accordingly, the initial state of the Markov chain is reachable from any states of the system. Conversely, using the same approach, it is easy to see that any states of the system is reachable from the initial state, $\mathbold{Z}(0)$. Consequently, the Markov chain $\mathbold{Z}(t)$ is irreducible. \\
Aperiodic: Since Markov chain $\mathbold{Z}(t)$ is irreducible, in order to show that it is also aperiodic, it suffices to show that there is a positive probability for transition from a state to itself. Due to the fact that there is a positive probability that zero task arrives to the system, the Markov chain stays at the initial state with a positive probability. Hence, the Markov chain $\mathbold{Z}(t)$ is aperiodic.


\subsection{Proof of Lemma \ref{newlemma2}}
Lemma \ref{newlemma2}:
For any arrival rate vector inside the capacity region, $\mathbold{\lambda} \in \Lambda$, there exists a load decomposition $\{\lambda_{i, m}\}$ and $\delta > 0$ such that
\begin{equation*}
\sum_{i \in \mathcal{L}} \frac{\lambda_{i, m}}{\mu_{i, m}} < \frac{1}{1 + \delta}, \ \forall m \in \mathcal{M}.
\end{equation*}

\textbf{Proof:}
The capacity region $\Lambda$ is an open set, so for any $\mathbold{\lambda} \in \Lambda$, there exists $\delta > 0$ such that $(1 + \delta) \mathbold{\lambda} = \mathbold{\lambda}' \in \Lambda$. On that account, \eqref{capacityregion} follows by $\sum_{i \in \mathcal{L}} \frac{\lambda_{i, m}'}{\mu_{i, m}} = \sum_{i \in \mathcal{L}} \frac{(1 + \delta )\lambda_{i, m}}{\mu_{i, m}} < 1, \forall m \in \mathcal{M}$, which completes the proof:
\[
\sum_{i \in \mathcal{L}} \frac{\lambda_{i, m}}{\mu_{i, m}} < \frac{1}{1 + \delta}, \forall m \in \mathcal{M}.
\]


\subsection{Proof of Lemma \ref{lemma123}}
\label{prooflemma123}
Lemma \ref{lemma123}:
\begin{equation*}
\langle \mathbold{W}(t) , \widetilde{\mathbold{U}}(t) \rangle = 0, \ \forall t.
\end{equation*}

\textbf{Proof:}
\[
\langle \mathbold{W}(t) , \widetilde{\mathbold{U}}(t) \rangle = \hspace{-0.2cm} \sum_{m \in \mathcal{M}} \hspace{-0.2cm} \bigg ( \frac{Q_m^1(t)}{\alpha_m^1} + \frac{Q_m^2(t)}{\alpha_m^2} + \cdots + \frac{Q_m^N(t)}{\alpha_m^{N^m}} \bigg ) \frac{U_m(t)}{\alpha_m^{N^m}}.
\]
If the unused service for server $m$ is zero, $U_m(t) = 0$, the corresponding term for server $m$ is zero in the above summation. Alternatively, the unused service of server $m$ is positive if and only if all $N^m$ sub-queues of the server are empty, which again makes the corresponding term for server $m$ in the above summation equal to zero.

\subsection{Proof of Lemma \ref{lemma1234}}
\label{prooflemma1234}
Lemma \ref{lemma1234}:
Under the exploration-exploitation routing policy of the Blind GB-PANDAS algorithm, for any arrival rate vector inside the capacity region, $\mathbold{\lambda} \in \Lambda$, and the corresponding ideal workload vector $\mathbold{w}$ defined in  \eqref{workloadm}, and for any arbitrary small $\theta_0 > 0$, there exists $T_0 > t_0$ such that for any $t_0 \geq 0$ and $T > T_0$:
\begin{equation*}
\begin{aligned}
 & \mathbb{E} \Big [ \sum_{t = T_0}^{t_0 + T - 1} \Big ( \langle \mathbold{W}(t), \mathbold{A}(t) \rangle - \langle \mathbold{W}(t), \mathbold{w} \rangle \Big ) \Big | \mathbold{Z}(t_0) \Big ] \\
 \leq & \theta_0 T || \mathbold{Q}(t_0) ||_1 + c_0,
\end{aligned}
\end{equation*}
where the constants $\theta_0, c_0 > 0$ are independent of $\mathbold{Z}(t_0)$.

\textbf{Proof:}
By the choice of exploration rate 
 for Blind GB-PANDAS, which is independent of the system state, and the fact that exploration exists in both routing and scheduling, any task that is $n$-local to server $m$ is scheduled on this server for infinitely many times in the interval $[t_0, \infty)$ only due to exploration, regardless of the initial system state.
 Processing time of an $n$-local task on server $m$ has
  a finite mean. Hence, due to strong law of large numbers, using the update rule \eqref{estimators} for the elements of the service rate matrix, we have:
\begin{equation}
\begin{aligned}
\label{epsilonchoice}
& \forall \ 0 < \epsilon < \frac12 \times \min \big \{ \min_{n \neq n', m} \big |\alpha_m^n - \alpha_m^{n'} \big |, \min_{m, n} \alpha_m^n, 0.5 \big \} \\
& \text{ and } \forall \delta' > 0, \exists \hspace{0.35mm} T_0' > t_0, \text{ such that for any } \mathbold{Z}(t_0) \\
& P \bigg ( \big | \widetilde{\alpha}_m^n(t) - \alpha_m^n \big | \hspace{-0.75mm} < \hspace{-0.5mm} \epsilon, \ 1 \hspace{-0.5mm} - \hspace{-0.5mm} \epsilon \hspace{-0.5mm} < \hspace{-0.5mm} \frac{\alpha_m^n}{\widetilde{\alpha}_m^n(t)} \hspace{-0.5mm} < \hspace{-0.5mm} 1 \hspace{-0.5mm} + \hspace{-0.5mm} \epsilon  \bigg | \mathbold{Z}(t_0) \bigg ) \hspace{-0.5mm} > \hspace{-0.5mm} 1 - \delta', \\
& \ \ \ \ \ \ \ \ \ \ \ \ \ \ \ \ \ \ \ \ \ \ \ \ \ \ \ \ \forall t > T_0', \ \forall m \in \mathcal{M}, \ \forall n \in \{1, 2, \cdots, N^m\}.
\end{aligned}
\end{equation}
By the above choice of $\epsilon$, for $t > T_0'$, the different locality levels are distinct from each other with at least $1 - \delta'$ probability. Let $E$ be the event that $\big | \widetilde{\alpha}_m^n(t) - \alpha_m^n \big | < \epsilon$ and $\ 1 - \epsilon < \frac{\alpha_m^n}{\widetilde{\alpha}_m^n(t)} < 1 + \epsilon$ for $t \geq T_0'$.

For an incoming task of type $i \in \mathcal{L}$ at time slot $t$, define the \textit{exact} (but not known) and \textit{estimated} minimum weighted workloads as
\begin{equation}
\begin{aligned}
\label{WWW}
& \overline{W}_i^*(t) = \min_{m \in \mathcal{M}} \frac{W_m(t)}{\mu_{i, m}}, \ \ \ \ \ \ \  \widetilde{\overline{W}}_i^*(t) = \min_{m \in \mathcal{M}} \frac{\widetilde{W}_m(t)}{\widetilde{\mu}_{i, m}(t)},
\end{aligned}
\end{equation}
where $W_m(t)$ and $\widetilde{W}_m(t)$ are defined in \eqref{exactWW} and \eqref{estimatedWW}, respectively.
$W_m(t)$ and $\widetilde{W}_m(t)$ are related to each other as follows:
\[
\begin{aligned}
\widetilde{W}_m(t) & = \frac{Q_m^1(t)}{\widetilde{\alpha}_m^1(t)} + \frac{Q_m^2(t)}{\widetilde{\alpha}_m^2(t)} + \cdots + \frac{Q_m^{N^m}(t)}{\widetilde{\alpha}_m^{N^m}(t)} \\
& = \frac{\alpha_m^1}{\widetilde{\alpha}_m^1(t)} \cdot \frac{Q_m^1(t)}{\alpha_m^1} + \cdots + \frac{\alpha_m^{N^m}}{\widetilde{\alpha}_m^{N^m}(t)} \cdot \frac{Q_m^{N^m}(t)}{\alpha_m^{N^m}},
\end{aligned}
\]
hence, using \eqref{epsilonchoice}, for any $t > T_0'$ and any $m \in \mathcal{M}$, we have
\begin{equation}
\label{WWWW}
P \bigg ( (1 - \epsilon)W_m(t) < \widetilde{W}_m(t) < (1 + \epsilon)W_m(t) \bigg | \mathbold{Z}(t_0), E \bigg ) = 1,
\end{equation}
and using \eqref{WWW} and \eqref{WWWW}, we have
\begin{equation}
\label{WWWWW}
P \bigg ( \frac{W_m(t)}{\mu_{i, m}} \geq \overline{W}_i^*(t) > \frac{1}{(1 + \epsilon)^2} \widetilde{\overline{W}}_i^*(t)  \bigg | \mathbold{Z}(t_0), E \bigg )  = 1.
\end{equation}

Using the conditional independence of $\widetilde{\mathbold{W}}(t)$ and $\mathbold{A}(t)$ from $\mathbold{Z}(t_0)$ given $\mathbold{Z}(t)$, for any $T > T_0' - t_0$, we have the following for $T_0' \leq t \leq t_0 + T - 1$:
\[
\begin{aligned}
& \mathbb{E} \big [ \langle \mathbold{W}(t), \mathbold{A}(t) \rangle | \mathbold{Z}(t_0) \big ] \\
\overset{(a)}{=} & \ \hspace{-0.5mm} \mathbb{E} \bigg [ \sum_{m \in \mathcal{M}} \hspace{-1.3mm} W_m(t) \Big ( \hspace{-0.5mm} \frac{A_m^1(t)}{\alpha_m^1} \hspace{-0.5mm} + \hspace{-0.5mm} \frac{A_m^2(t)}{\alpha_m^2} \hspace{-0.5mm} + \hspace{-0.5mm} \cdots \hspace{-0.5mm} + \hspace{-0.5mm} \frac{A_m^{N^m}(t)}{\alpha_m^{N^m}} \hspace{-0.25mm} \Big ) \hspace{-0.5mm} \bigg | \mathbold{Z}(t_0) \hspace{-0.5mm} \bigg ] \\
\overset{(b)}{=} & \ \hspace{-0.04cm} \mathbb{E} \bigg [ \hspace{-0.04cm} \sum_m W_m(t) \bigg ( \frac{1}{\alpha_m^1} \sum_{\substack{i \in \mathcal{L}_m^1}} \hspace{-0.1cm} A_{i, m}(t) + \frac{1}{\alpha_m^2} \sum_{\substack{i \in \mathcal{L}_m^2}} \hspace{-0.1cm} A_{i, m}(t) + \\
& \ \ \ \ \ \ \ \ \ \ \ \ \ \ \ \ \ \ \ \ \ \ \ \ \ \ \ \ \ \ \ \ \ \ \hspace{0.61mm} \cdots + \frac{1}{\alpha_m^{N^m}} \sum_{\substack{i \in \mathcal{L}_m^{N^m}}} A_{i, m}(t) \bigg ) \bigg | \mathbold{Z}(t_0) \bigg ] \\
\overset{(c)}{=} & \ \mathbb{E} \bigg [ \sum_{i \in \mathcal{L}} \sum_{m \in \mathcal{M}} \bigg ( \frac{W_m(t)}{\mu_{i, m}} A_{i, m}(t) \bigg ) \bigg | \mathbold{Z}(t_0) \bigg ] \\
\overset{(d)}{\leq} & \ \mathbb{E} \bigg [ \sum_{i \in \mathcal{L}} \sum_{m \in \mathcal{M}} \bigg ( \frac{1}{(1 - \epsilon)^2} \cdot \frac{\widetilde{W}_m(t)}{\widetilde{\mu}_{i, m}} A_{i, m}(t) \bigg ) \bigg | \mathbold{Z}(t_0), E \bigg ] \\
& \hspace{-4mm} + \delta' \hspace{-1mm} \cdot \hspace{-0.5mm} \mathbb{E} \bigg [ \sum_{i \in \mathcal{L}} \sum_{m \in \mathcal{M}} \hspace{-1.6mm} \bigg ( \frac{Q_m^1(t) \hspace{-0.4mm} + \cdots + \hspace{-0.4mm} Q_m^{N^m} \hspace{-0.4mm} (t)}{ \underset{i, m}{\min} \{ \mu_{i, m} \} \hspace{-0.7mm} \cdot \hspace{-0.6mm} \underset{i}{\min} \{ \mu_{i, m} \}} A_{i, m}(t) \hspace{-0.5mm} \bigg ) \hspace{-0.5mm} \bigg | \mathbold{Z}(t_0), E^c \bigg ]
\end{aligned}
\]
\begin{equation}
\begin{aligned}
& \overset{(e)}{<} \mathbb{E} \bigg [ \mathbb{E} \bigg [ \sum_{i \in \mathcal{L}} \bigg ( p_e \cdot \frac{1}{(1 - \epsilon)^2} \cdot \widetilde{\overline{W}}_i^*(t) A_i(t) +  \frac{ 1 - p_e }{( 1  - \epsilon )^2} \times \\
& \sum_m \hspace{-0.05cm} \frac{ \sum_{n = 1}^{N^m} Q_m^n(t_0) 
  + \hspace{-0.05cm} N_T (T \hspace{-0.05cm} - \hspace{-0.04cm} t_0 \hspace{-0.04cm} )C_A}{ \underset{i, m}{\min} \{ \widetilde{\mu}_{i, m}(t) \} \cdot \underset{i}{\min} \{ \widetilde{\mu}_{i, m}(t) \}} \hspace{-0.05cm} \cdot \hspace{-0.05cm} C_A \bigg ) \bigg | \mathbold{Z}(t) \bigg ] \bigg | \mathbold{Z}(t_0) , E \bigg ] + \\
& \delta' \hspace{-1mm} \cdot \hspace{-0.5mm} \mathbb{E} \hspace{-0.4mm} \bigg [ \hspace{-0.5mm}  \sum_{i \in \mathcal{L}} \hspace{-0.7mm} \bigg ( \hspace{-0.99mm} \sum_m \hspace{-0.6mm} \frac{ \sum_{n = 1}^{N^m} \hspace{-0.5mm} Q_m^n \hspace{-0.1mm} (t_0) \hspace{-0.5mm} + \hspace{-0.5mm} N_T (T \hspace{-0.5mm} - t_0)C_A}{  \min_{i, m} \{ \mu_{i, m} \} \cdot \min_i \{ \mu_{i, m} \}} \hspace{-0.39mm} \cdot \hspace{-0.39mm} C_A \hspace{-0.99mm} \bigg ) \hspace{-0.7mm} \bigg | \mathbold{Z}(t_0), E^c \bigg ] \\
& \hspace{-0.5mm} \overset{(f)}{<} \hspace{-1.5mm} \frac{1}{( \hspace{-0.4mm} 1 \hspace{-1.1mm} - \hspace{-0.7mm} \epsilon \hspace{-0.2mm} )^2} \hspace{-0.75mm} \sum_{i \in \mathcal{L}} \hspace{-0.3mm} \mathbb{E} \hspace{-0.4mm} \bigg [ \hspace{-0.4mm} \widetilde{\overline{W}}_i^* \hspace{-0.6mm} (t) \hspace{-0.3mm}  \bigg | \mathbold{Z}(t_0), E \hspace{-0.3mm}  \bigg ] \hspace{-0.5mm} \lambda_i \hspace{-0.75mm} + \hspace{-1.1mm} \left ( \hspace{-0.9mm} \frac{1}{t^{\delta''}} \hspace{-0.8mm} + \hspace{-0.5mm} \delta' \hspace{-1.1mm} \right ) \hspace{-0.75mm} c_0'' \| \mathbold{Q}(t_0) \|_1 \hspace{-1mm} + \hspace{-0.7mm} c_0',
\label{eq1}
\end{aligned}
\end{equation}
where $(a)$ and $(b)$ are simply followed by the definitions of pseudo task arrival process in \eqref{pseudoparameters} and $A_m^n(t)$ in \eqref{sub-queue-arrival}, respectively. The order of summations is changed in $(c)$. By the law of total probability, \eqref{epsilonchoice}, and \eqref{WWWW}, $(d)$ is true, and $(e)$ follows by the routing policy of Blind GB-PANDAS, where an incoming task at the beginning of time slot $t$ is routed to the corresponding sub-queue of the server with the minimum estimated weighted workload with probability $p_e = \max (1 - p(t), 0)$ 
 and is routed to the corresponding sub-queue of a server chosen uniformly at random with probability $1 - p_e$. 
  Also note that the number of arriving tasks at a time slot is assumed to be upper bounded by $C_A$. The last step, $(f)$, is true by using \eqref{epsilonchoice}, upper bounding the exploration probability $1 - p_e$ by $\frac{1}{t^{\delta''}}$ given that $\delta'' > 0$ is a constant, and doing simple calculations, where $c_0''$ and $c_0'$ are constants independent of $\mathbold{Z}(t_0)$. Note that minimum value of the estimated service rates, $\min_{i, m} \{ \widetilde{\mu}_{i, m}(t) \}$, is lower bounded for any $t \geq t_0$ by a constant which is the minimum of the initialization of service rates and the inverse of the maximum support of CDF functions $F_{i, m}$.
We also have
\begin{equation}
\begin{aligned}
& \mathbb{E} \big [ \langle \mathbold{W} (t), \mathbold{w} \rangle | \mathbold{Z}(t_{0}) \big ] = \mathbb{E} \left [ \sum_{m \in \mathcal{M}} W_m(t) w_m \bigg | \mathbold{Z}(t_{0}) \right ] \overset{(a)}{=} \\
& \mathbb{E} \left [ \sum_{m \in \mathcal{M}} \hspace{-2.5mm} \Big ( \hspace{-0.5mm} W_{ \hspace{-0.2mm} m} \hspace{-0.2mm} (t) \hspace{-1.0mm} \sum_{i \in \mathcal{L}} \hspace{-0.6mm} \frac{\lambda_{i, m}}{\mu_{i, m}} \hspace{-0.5mm} \Big ) \hspace{-0.3mm}  \bigg | \hspace{-0.2mm} \mathbold{Z}(t_{0}) \hspace{-0.5mm} \right ] \hspace{-1.35mm}
\overset{(b)}{=} \hspace{-0.65mm} \mathbb{E} \hspace{-1.1mm} \Bigg [ \sum_{\underset{m \in \mathcal{M}}{i \in \mathcal{L}}  } \hspace{-1.5mm} \frac{W_{\hspace{-0.2mm} m} \hspace{-0.2mm} (t)}{\mu_{i, m}} \lambda_{i, m} \hspace{-0.1mm} \bigg | \hspace{-0.2mm} \mathbold{Z}(t_{0}) \hspace{-0.8mm} \Bigg ] \hspace{-1.5mm} \overset{(c)}{\geq} \\
& \hspace{-1.5mm} \sum_{\underset{m \in \mathcal{M}}{i \in \mathcal{L}}} \hspace{-1.5mm} \frac{1 \hspace{-0.2mm} - \hspace{-0.2mm} \delta'}{( \hspace{-0.5mm} 1 \hspace{-1mm} + \hspace{-0.9mm} \epsilon \hspace{-0.4mm} )^2} \hspace{-0.2mm}  \mathbb{E \hspace{-0.6mm} \bigg [ } \hspace{-0.4mm} \widetilde{\overline{W}}_i^* \hspace{-0.5mm} ( \hspace{-0.3mm} t \hspace{-0.4mm} ) \hspace{-0.3mm}  \bigg | \hspace{-0.3mm} \mathbold{Z}( \hspace{-0.3mm} t_0 \hspace{-0.5mm} ), \hspace{-0.5mm} E \hspace{-0.3mm}  \bigg ] \hspace{-0.55mm} \lambda_{i, m} \hspace{-1mm} = \hspace{-1mm} \frac{1 \hspace{-0.2mm} - \hspace{-0.2mm} \delta'}{( \hspace{-0.5mm} 1 \hspace{-1mm} + \hspace{-0.9mm} \epsilon \hspace{-0.4mm} )^2} \hspace{-0.95mm} \sum_{i \in \mathcal{L}} \hspace{-0.5mm} \mathbb{E} \hspace{-0.5mm} \bigg [ \hspace{-0.5mm}  \widetilde{\overline{W}}_i^* \hspace{-0.75mm} ( \hspace{-0.3mm} t \hspace{-0.4mm} ) \hspace{-0.5mm}  \bigg | \hspace{-0.4mm} \mathbold{Z}( \hspace{-0.3mm} t_0 \hspace{-0.5mm} ), \hspace{-0.5mm} E \hspace{-0.5mm}  \bigg ] \hspace{-0.6mm} \lambda_i,
\label{eq2}
\end{aligned}
\end{equation}
where $(a)$ is true by the definition of the ideal workload on a server in \eqref{workloadm}, note that the ideal workload is not state dependent but $W_m(t)$ is, the order of summations is changed in $(b)$, and $(c)$ is followed by the law of total probability, ignoring the second term, and Equation \eqref{WWWWW}.

Putting \eqref{eq1} and \eqref{eq2} together, for $T > T_0 > T_0'$, we have
\[
\begin{aligned}
& \mathbb{E} \Big [ \sum_{t = T_0}^{t_0 + T - 1} \Big ( \langle \mathbold{W}(t), \mathbold{A}(t) \rangle - \langle \mathbold{W}(t), \mathbold{w} \rangle \Big ) \Big | \mathbold{Z}(t_0) \Big ] \\
< & \hspace{-0.5mm} \sum_{t = T_0}^{t_0 + T - 1} \hspace{-1mm} \Bigg ( \hspace{-1.5mm} \left ( \hspace{-0.5mm} \frac{1}{(1 - \epsilon)^2} \hspace{-0.25mm} - \hspace{-0.25mm} \frac{1 - \delta'}{(1 + \epsilon)^2} \hspace{-0.5mm} \right ) \hspace{-0.8mm} \sum_{i \in \mathcal{L}} \mathbb{E} \bigg [ \widetilde{\overline{W}}_i^*(t)  \bigg | \mathbold{Z}(t_0), E  \bigg ] \lambda_i \\
& + \left ( \frac{1}{t^{\delta''}} + \delta' \right )c_0'' \| \mathbold{Q}(t_0) \|_1 + c_0' \Bigg )
\end{aligned}
\]
\[
\begin{aligned}
\overset{(a)}{<} & \frac{16}{9}  \left ( 4\epsilon + \delta'  \right )  \cdot \Bigg (  \sum_{t = T_0}^{t_0 + T - 1}  \sum_{i \in \mathcal{L}} \hspace{0.1cm} \mathbb{E} \bigg [ \widetilde{\overline{W}}_i^*(t) \bigg | \mathbold{Z}(t_0), E \bigg ] \lambda_i  \Bigg ) \\
& + T  \left (  \frac{1}{T_0^{\delta''}}  +  \delta'  \right )  c_0'' \| \mathbold{Q}(t_0) \|_1  +  T c_0' \\
\overset{(b)}{<} & \frac{16}{9} \left ( 4\epsilon + \delta' \right ) T N_T \max_{i} \{\lambda_i\} \\
& \times \left ( \mathbb{E} \bigg [ \sum_m \frac{ \sum_{n = 1}^{N^m} Q_m^1(t_0)  + N_T (T - t_0)C_A}{\min_{i, m} \{ \widetilde{\mu}_{i, m}(t) \} \cdot \min_i \{ \widetilde{\mu}_{i, m}(t) \}}  \bigg | \mathbold{Z}(t_0), E \bigg ] \right ) \\
& + T \left ( \frac{1}{T_0^{\delta''}} + \delta' \right )c_0'' \| \mathbold{Q}(t_0) \|_1 + T c_0' \\
\overset{(c)}{<} & \left ( \epsilon + \delta' + \frac{1}{T_0^{\delta''}} \right ) T c_1 \| \mathbold{Q}(t_0) \|_1 + c_0 = \theta_0 T \| \mathbold{Q}(t_0) \|_1 + c_0,
\end{aligned}
\]
where $(a)$ follows by upper bounding $1 - \epsilon$, $\frac{1}{(1 - \epsilon)^2 (1 + \epsilon)^2}$, and $\frac{1}{t^{\delta''}}$ by $1$, $\frac{16}{9}$, and $\frac{1}{T_0^{\delta''}}$, respectively, and $(b)$ is true by the fact that the number of arriving tasks is bounded by $C_A$, the number of task types is $N_T$, and the maximum arrival rate of task types, $\max_i \{\lambda_i\}$, is bounded by the number of servers. Inequality $(c)$ is true by doing simple calculations and using the fact that $\min_{i, m} \{ \widetilde{\mu}_{i, m}(t) \}$ is lower bounded by a constant for any $t \geq t_0$ as discussed in $(f)$ of \eqref{eq1}. \\
\textbf{Remark.} $\theta_0$ can be made arbitrary small by choosing $\epsilon$ and $\delta'$ small and $T_0$ large enough.

\subsection{Proof of Lemma \ref{lemma12345}}
\label{prooflemma12345}
Lemma \ref{lemma12345}:
Under the exploration-exploitation scheduling policy of the Blind GB-PANDAS algorithm, for any arrival rate vector inside the capacity region, $\mathbold{\lambda} \in \Lambda$, and the corresponding ideal workload vector $\mathbold{w}$ in \eqref{workloadm}, there exists $T_1 > 0$ such that for any $T > T_1$, we have:
\begin{equation*}
\begin{aligned}
 & \mathbb{E} \left [ \sum_{t = t_0}^{t_0 + T - 1} \Big ( \langle \mathbold{W}(t), \mathbold{w} \rangle - \langle \mathbold{W}(t), \mathbold{S}(t) \rangle \Big ) \Big | \mathbold{Z}(t_0) \right ] \\
 \leq & - \theta_1 T || {\mathbold{Q}}(t_0) ||_1 + c_1, \ \forall t_0 \geq 0,
\end{aligned}
\end{equation*}
where the constants $\theta_1, c_1 > 0$ are independent of $\mathbold{Z}(t_0)$.

\textbf{Proof:}
The proof is similar to the proof of lemma 4 in \cite{yekkehkhany2017gb} and is presented for the sake of completeness. By the assumption on boundedness of arrival and service processes, there exists a constant $C_A$ such that for any $t_0, t,$ and $T$ with $t_0 \leq t \leq t_0 + T$, we have the following for all $m \in \mathcal{M}$:
\begin{equation}
W_m(t_0) - \frac{T}{\min_n \{\alpha_m^n\}} \leq W_m(t) \leq W_m(t_0) + \frac{T C_A}{\min_n \{\alpha_m^n\}}.
\label{boundT}
\end{equation}
On the other hand, by Lemma \ref{newlemma2}, the ideal workload on a server defined in \eqref{workloadm} can be bounded as follows:
\begin{equation}
w_m \leq \frac{1}{1 + \delta}, \ \forall m \in \mathcal{M}.
\label{workloadbound}
\end{equation}
Hence,
\begin{equation*}
\begin{aligned}
& \mathbb{E} \left [ \sum_{t = t_0}^{t_0 + T - 1} \Big ( \langle \mathbold{W}(t), \mathbold{w} \rangle \Big ) \Big | \mathbold{Z}(t_0) \right ] \\
= \ & \mathbb{E} \left [ \sum_{t = t_0}^{t_0 + T - 1} \left ( \sum_{m = 1}^M W_m(t) w_m \right ) \bigg | \mathbold{Z}(t_0) \right ]
\end{aligned}
\end{equation*}
\begin{equation}
\begin{aligned}
\overset{(a)}{\leq} & \ T \sum_{m = 1}^M \Big ( W_m(t_0) w_m  + \frac{MT^2C_A}{\min_n \{ \alpha_m^n \}} \Big ) \\
\overset{(b)}{\leq} & \ \frac{T}{1 + \delta} \sum_m W_m(t_0) + \frac{MT^2C_A}{\min_{m, n} \{ \alpha_m^n \}},
\label{eq21}
\end{aligned}
\end{equation}
where $(a)$ is true by bringing the inner summation on $m$ out of the expectation and using the boundedness property of the workload in Equation \eqref{boundT}, and $(b)$ is true by Equation \eqref{workloadbound}.

Before investigating the second term on the left-hand side of Equation \eqref{IIII}, $\mathbb{E} \Big [$ $\sum_{t = t_0}^{t_0 + T - 1}$ $\Big ( \langle \mathbold{W}(t), \mathbold{S}(t) \rangle \Big )$ $\Big | \mathbold{Z}(t_0)  \Big ]$, we propose the following lemma which will be used in lower bounding this second term.

\begin{lemma}
For any server $m \in \mathcal{M}$ and any $t_0$, we have the following:
\[
\hspace{-0.5cm} \lim_{T \rightarrow \infty} \hspace{-1mm} \frac{ \mathbb{E} \hspace{-0.6mm} \left [ \hspace{-0.1mm} \sum_{t = t_0}^{t_0 + T - 1} \hspace{-1mm} \bigg ( \hspace{-0.5mm} \frac{S_m^1(t)}{\alpha_m^1} \hspace{-0.5mm} + \hspace{-0.5mm} \frac{S_m^2(t)}{\alpha_m^2} \hspace{-0.5mm} + \hspace{-0.5mm} \cdots \hspace{-0.5mm} + \hspace{-0.5mm} \frac{S_m^{N^m}(t)}{\alpha_m^{N^m}} \hspace{-0.5mm} \bigg ) \bigg | \mathbold{Z}(t_0) \right ]}{T} = 1.
\]
\label{asymptoticT}
\end{lemma}
We then have the following:
\begin{equation}
\label{eq22}
\begin{aligned}
& \mathbb{E} \left [ \sum_{t = t_0}^{t_0 + T - 1} \Big ( \langle \mathbold{W}(t), \mathbold{S}(t) \rangle \Big ) \Big | \mathbold{Z}(t_0)  \right ] \\
= \ & \mathbb{E} \Bigg [ \sum_{t = t_0}^{t_0 + T - 1} \sum_{m = 1}^{M} \Bigg ( W_m(t) \bigg ( \frac{S_m^1(t)}{\alpha_m^1} + \frac{S_m^2(t)}{\alpha_m^2} + \\
& \ \ \ \ \ \ \ \ \ \ \ \ \ \ \ \ \ \ \ \ \ \ \ \ \ \ \ \ \ \ \ \ \ \ \ \ \ \ \ \ \ \ \ \ \ \ \ \ \ \ \cdots + \frac{S_m^{N^m}(t)}{\alpha_m^{N^m}} \bigg ) \Bigg ) \bigg | \mathbold{Z}(t_0) \Bigg ] \\
\overset{(a)}{ \geq } & \ \sum_{m = 1}^{M} \Bigg ( W_m(t_0) \mathbb{E} \Bigg [ \sum_{t = t_0}^{t_0 + T - 1} \bigg ( \frac{S_m^1(t)}{\alpha_m^1} + \frac{S_m^2(t)}{\alpha_m^2} + \\
& \ \ \ \ \ \ \ \ \ \ \ \ \ \ \ \ \ \ \ \ \ \ \ \ \ \ \ \ \ \ \ \ \ \ \ \ \ \ \ \ \ \ \ \ \ \ \ \ \ \ \cdots + \frac{S_m^{N^m}(t)}{\alpha_m^{N^m}} \bigg ) \bigg | \mathbold{Z}(t_0) \Bigg ] \Bigg ) \\
& - \sum_{m = 1}^{M} \Bigg ( \frac{T}{\min_n \{ \alpha_m^n \}} \mathbb{E} \bigg [ \sum_{t = t_0}^{t_0 + T - 1} \bigg ( \frac{S_m^1(t)}{\alpha_m^1} + \frac{S_m^2(t)}{\alpha_m^2} + \\
& \ \ \ \ \ \ \ \ \ \ \ \ \ \ \ \ \ \ \ \ \ \ \ \ \ \ \  \ \ \ \ \ \ \ \ \ \ \ \ \ \ \ \ \ \ \ \ \ \cdots + \frac{S_m^{N^m}(t)}{\alpha_m^{N^m}} \bigg ) \bigg | \mathbold{Z}(t_0) \bigg ] \Bigg ), \\
\end{aligned}
\end{equation}
where $(a)$ follows by bringing the inner summation on $m$ out of the expectation and using the boundedness property of the workload in Equation \eqref{boundT}.

Using Lemma \ref{asymptoticT}, for any $0 < \epsilon_0 < \frac{\delta}{1 + \delta}$, there exists $T_1$ such that for any $T \geq T_1$, we have the following for any server $m \in \mathcal{M}$:
\[
\begin{aligned}
1 & - \epsilon_0 \leq \\
& \ \frac{ \mathbb{E} \left [ \sum_{t = t_0}^{t_0 + T - 1} \bigg ( \frac{S_m^1(t)}{\alpha_m^1} + \frac{S_m^2(t)}{\alpha_m^2} + \cdots + \frac{S_m^{N^m}(t)}{\alpha_m^{N^m}} \bigg ) \bigg | \mathbold{Z}(t_0) \right ]}{T} \\
&  \ \ \ \ \ \ \ \ \ \ \ \ \ \ \ \ \ \ \ \ \ \ \ \ \ \ \ \ \ \ \ \ \ \ \ \ \ \ \ \ \ \ \ \ \ \ \ \ \ \ \ \ \ \ \ \ \ \ \ \ \ \ \ \ \ \ \ \ \ \ \ \ \ \ \leq 1 + \epsilon_0.
\end{aligned}
\]
Then continuing on Equation \eqref{eq22}, we have the following:
\begin{equation}
\label{eq22222}
\begin{aligned}
& \mathbb{E} \left [ \sum_{t = t_0}^{t_0 + T - 1} \Big ( \langle \mathbold{W}(t), \mathbold{S}(t) \rangle \Big ) \Big | \mathbold{Z}(t_0)  \right ] \\
\geq & T (1 - \epsilon_0)\sum_{m = 1}^{M} W_m(t_0) - \frac{MT^2(1 + \epsilon_0)}{\min_{m, n} \{ \alpha_m^n \}}. \\
\end{aligned}
\end{equation}

Then Lemma \ref{lemma12345} is concluded as follows by using equations \eqref{eq21} and \eqref{eq22222} and picking $c_1 = \frac{M T^2}{\min_{m, n} \{ \alpha_m^n \}} (C_A + 1 + \epsilon_0)$ and $\theta_1 = \frac{1}{\max_{m, n} \{ \alpha_m^n \}} \left ( \frac{\delta}{1 + \delta} - \epsilon_0 \right )$, where by our choice of $\epsilon_0$ we have $\theta_1 > 0$:
\[
\begin{aligned}
& \mathbb{E} \left [ \sum_{t = t_0}^{t_0 + T - 1} \Big ( \langle \mathbold{W}(t), \mathbold{w} \rangle - \langle \mathbold{W}(t), \mathbold{S}(t) \rangle \Big ) \Big | \mathbold{Z}(t_0) \right ] \\
\leq & \hspace{-0.5mm} - \hspace{-0.5mm} T \hspace{-0.5mm} \left ( \hspace{-0.5mm} \frac{\delta}{1 \hspace{-0.5mm} + \hspace{-0.5mm} \delta} - \epsilon_0 \hspace{-0.75mm} \right ) \hspace{-0.9mm} \sum_{m = 1}^{M} \hspace{-0.5mm} W_m(t_0) +  \frac{M T^2}{\min_{m, n} \{ \alpha_m^n \}} (C_A \hspace{-0.5mm} + \hspace{-0.5mm} 1 \hspace{-0.5mm} + \hspace{-0.5mm} \epsilon_0)
\end{aligned}
\]
\[
\begin{aligned}
\overset{(a)}{\leq} & - \frac{T}{\max_{m, n} \{ \alpha_m^n \}} \left ( \frac{\delta}{1 + \delta} - \epsilon_0 \right ) \sum_{m = 1}^{M} \Big (Q_m^1(t_0) + Q_m^2(t_0) + \\
& \ \ \ \ \ \ \ \ \ \ \ \ \ \ \ \ \ \ \ \ \ \ \ \ \ \ \ \ \ \ \ \ \ \ \ \ \ \ \ \ \ \ \ \ \ \ \ \ \ \ \ \ \ \ \ \ \ \cdots + Q_m^{N^m}(t_0) \Big ) + c_1 \\
\leq & - \theta_1 T \| {\mathbold{Q}}(t_0) \|_1 + c_1, \ \forall T \geq T_0,
\end{aligned}
\]
where $(a)$ is true as $W_m(t_0) \geq \frac{Q_m^1(t_0) + Q_m^2(t_0) + \cdots + Q_m^{N^m}(t_0)}{\max_{m, n} \{ \alpha_m^n \}}$.

\subsection{Proof of Lemma \ref{lemma123456}}
\label{prooflemma123456}
Lemma \ref{lemma123456}:
Under the exploration-exploitation load balancing of the Blind GB-PANDAS algorithm, for any arrival rate vector inside the capacity region, $\mathbold{\lambda} \in \Lambda$, and for any $\theta_2 > 0$, there exists $T_2 > 0$ such that for any $T > T_2$ and for any $t_0 \geq 0$, we have:
\[
\mathbb{E} \Big [ ||\mathbold{\Psi}(t_0 + T)||_1 - ||\mathbold{\Psi}(t_0)||_1 \Big | \mathbold{Z}(t_0) \Big ] \leq -\theta_2 ||\mathbold{\Psi}(t_0)||_1 + M T.
\]

\textbf{Proof:}
This proof is the same as the proof of lemma 5 in \cite{yekkehkhany2017gb} and is presented for the sake of completeness. For any server $m \in \mathcal{M}$, let $t_m^*$ be the first time slot after or at time slot $t_0$ at which the server is available; that is,
\begin{equation}
\label{ali_t_star}
t_m^* = \min\{ \tau: \tau \geq t_0, \Psi_m(\tau) = 0 \},
\end{equation}
where it is obvious that $\Psi_m(t_m^*) = 0.$
Note that for any $t \geq t_0$, we have $\Psi_m(t + 1) \leq \Psi_m(t) + 1$, which is true by the definition of $\Psi(t)$ that is the number of time slots that server $m$ has spent on the currently in-service task. From time slot $t$ to $t+1$, if a new task comes in service, then $\Psi_m(t+1) = 0$ which results in $\Psi_m(t + 1) \leq \Psi_m(t) + 1$; otherwise, if server $m$ continues giving service to the same task, then $\Psi_m(t + 1) = \Psi_m(t) + 1$. Thus, if $t_m^* \leq t_0 + T$, it is easy to find out that $\Psi_m(t_0 + T) \leq t_0 + T - t_m^* \leq T$. In the following, we use $t_m^*$ to find a bound on $\mathbb{E} [ \Psi_m(t_0 + T) - \Psi_m(t_0) | \mathbold{Z}(t_0)]$:

\begin{equation}
\begin{aligned}
& \mathbb{E} \Big [ \|\mathbold{\Psi}(t_0 + T)\|_1 - \|\mathbold{\Psi}(t_0)\|_1 \Big | \mathbold{Z}(t_0) \Big ] \\
= & \sum_{m = 1}^M \mathbb{E} \left [ \Big (\Psi_m(t_0 + T) - \Psi_m(t_0) \Big ) \bigg | \mathbold{Z}(t_0) \right ] \\
= & \sum_{m = 1}^M \bigg \{ \mathbb{E} \left [ \Big (\Psi_m(t_0 + T) - \Psi_m(t_0) \Big ) \bigg | \mathbold{Z}(t_0), t_m^* \leq t_0 + T \right ] \\
& \times P \left ( t_m^* \leq t_0 + T \big | \mathbold{Z}(t_0) \right ) \\
& + \mathbb{E} \left [ \Big (\Psi_m(t_0 + T) - \Psi_m(t_0) \Big ) \bigg | \mathbold{Z}(t_0), t_m^* > t_0 + T \right ] \\
& \times P \left (t_m^* > t_0 + T \big | \mathbold{Z}(t_0) \right ) \bigg \} \\
\overset{(a)}{\leq} & \sum_{m = 1}^M \bigg \{ \Big (T - \Psi_m(t_0) \Big ) \times P \left (t_m^* > t_0 + T \big | \mathbold{Z}(t_0) \right ) \\
& + T \times P \left (t_m^* > t_0 + T \big | \mathbold{Z}(t_0) \right ) \bigg \} \\
= & - \sum_{m = 1}^M \bigg ( \Psi_m(t_0) \cdot P \left (t_m^* > t_0 + T \big | \mathbold{Z}(t_0) \right ) \bigg ) + MT,
\label{sdfg}
\end{aligned}
\end{equation}
where $(a)$ is true as given that $t_m^* \leq t_0 + T$ we found that $\Psi_m(t_0 + T) \leq T$, so $\Psi_m(t_0 + T) - \Psi_m(t_0) \leq T - \Psi_m(t_0)$, and given that $t_m^* > t_0 + T$, it is concluded that server $m$ is giving service to the same task over the whole interval $[t_0, t_0 + T]$, which results in $\Psi_m(t_0 + T) - \Psi_m(t_0) = T$.

Since the CDF of service time of an $n$-local task on server $m$
has finite mean, we have the following:
\[
\lim_{T \rightarrow \infty} P \left (t_m^* \leq t_0 + T \Big | \mathbold{Z}(t_0) \right ) = 1, \ \forall m \in \mathcal{M},
\]
so for any $\theta_2 \in (0, 1)$ there exists $T_2$ such that for any $T \geq T_2$, we have $P \left (t_m^* \leq t_0 + T \Big | \mathbold{Z}(t_0) \right ) \geq \theta_2,$ for any $m \in \mathcal{M}$, so Equation \eqref{sdfg} follows as below which completes the proof:
\begin{equation}
\begin{aligned}
& \mathbb{E} \Big [ \|\mathbold{\Psi}(t_0 + T)\|_1 - \|\mathbold{\Psi}(t_0)\|_1 \Big | \mathbold{Z}(t_0) \Big ] \\
\leq & - \theta_2 \sum_{m = 1}^M \Psi_m(t_0) + MT
= -\theta_2 \|\mathbold{\Psi}(t_0)\|_1 + M T.
\end{aligned}
\end{equation}



\subsection{Proof of Lemma \ref{lemma_drift}}
\label{prooflemma_drift}
Lemma \ref{lemma_drift}:
For any $t_0 \leq T_0 < T$, specifically $T_0$ from lemma \ref{lemma1234} that is dictated by choosing $\theta_0 < \theta_1$, we have the following for the drift of the Lyapunov function in \eqref{Lyapunov_function_1}, where $T_0$ is used in the first summation after the inequality:
\begin{equation*}
\begin{aligned}
 & \mathbb{E} \Big [ V(\mathbold{Z}(t_0 + T)) - V(\mathbold{Z}(t_0)) \Big | \mathbold{Z}(t_0) \Big ] \\
\leq & 2 \mathbb{E} \left [ \sum_{t = T_0}^{t_0 + T - 1} \Big ( \langle \mathbold{W}(t), \mathbold{A}(t) \rangle - \langle \mathbold{W}(t), \mathbold{w} \rangle \Big ) \Big | \mathbold{Z}(t_0) \right ] \\
 & + 2 \mathbb{E} \left [ \sum_{t = t_0}^{t_0 + T - 1} \Big ( \langle \mathbold{W}(t), \mathbold{w} \rangle - \langle \mathbold{W}(t), \mathbold{S}(t) \rangle \Big ) \Big | \mathbold{Z}(t_0) \right ] \\
 & + \mathbb{E} \Big [ || \mathbold{\Psi}(t_0 + T) ||_1 - || \mathbold{\Psi}(t) ||_1 \Big | \mathbold{Z}(t_0) \Big ] + c_2 \| \mathbold{Q}(t_0) \|_1 + c_3.
\end{aligned}
\end{equation*}

\textbf{Proof:}
\begin{equation}
\begin{aligned}
& \mathbb{E} \Big [ V(\mathbold{Z}(t_0 + T)) - V(\mathbold{Z}(t_0)) \Big | \mathbold{Z}(t_0) \Big ] \\
= & \mathbb{E} \Big [ \| \mathbold{W}(t_0 + T) \|^2 - \| \mathbold{W}(t_0) \|^2 \Big | \mathbold{Z}(t_0) \Big ] \\
& + \mathbb{E} \Big [ \| \mathbold{\Psi}(t_0 + T) \|_1 - \| \mathbold{\Psi}(t_0) \|_1 \Big | \mathbold{Z}(t_0) \Big ] \\
\overset{(a)}{=} & \mathbb{E} \left [ \sum_{t = t_0}^{t_0 + T - 1} \Big ( \| \mathbold{W}(t + 1) \|^2 - \| \mathbold{W}(t) \|^2 \Big ) \Big | \mathbold{Z}(t_0) \right ] \\
& + \mathbb{E} \Big [ \| \mathbold{\Psi}(t_0 + T) \|_1 - \| \mathbold{\Psi}(t) \|_1 \Big | \mathbold{Z}(t_0) \Big ] \\
\overset{(b)}{=} & \mathbb{E} \Bigg [ \sum_{t = t_0}^{t_0 + T - 1} \Big ( \| \mathbold{A}(t) - \mathbold{S}(t) +\widetilde{\mathbold{U}}(t)\|^2 \\
& + 2 \langle \mathbold{W}(t), \mathbold{A}(t) - \mathbold{S}(t) \rangle + 2 \langle \mathbold{W}(t), \widetilde{\mathbold{U}}(t) \rangle \Big ) \Big | \mathbold{Z}(t_0) \Bigg ] \\
& + \mathbb{E} \Big [ \| \mathbold{\Psi}(t_0 + T) \|_1 - \| \mathbold{\Psi}(t) \|_1 \Big | \mathbold{Z}(t_0) \Big ] \\
\overset{(c)}{\leq} & 2 \mathbb{E} \left [ \sum_{t = t_0}^{t_0 + T - 1} \Big ( \langle \mathbold{W}(t), \mathbold{A}(t) - \mathbold{S}(t) \rangle \Big ) \Big | \mathbold{Z}(t_0) \right ] \\
& + \mathbb{E} \Big [ \| \mathbold{\Psi}(t_0 + T) \|_1 - \| \mathbold{\Psi}(t) \|_1 \Big | \mathbold{Z}(t_0) \Big ] + c_3' \\
\overset{(d)}{=} \ & 2 \mathbb{E} \left [ \sum_{t = t_0}^{t_0 + T - 1} \Big ( \langle \mathbold{W}(t), \mathbold{A}(t) \rangle - \langle \mathbold{W}(t), \mathbold{w} \rangle \Big ) \Big | \mathbold{Z}(t_0) \right ] \\
& + 2 \mathbb{E} \left [ \sum_{t = t_0}^{t_0 + T - 1} \Big ( \langle \mathbold{W}(t), \mathbold{w} \rangle - \langle \mathbold{W}(t), \mathbold{S}(t) \rangle \Big ) \Big | \mathbold{Z}(t_0) \right ] \\
& + \mathbb{E} \Big [ \| \mathbold{\Psi}(t_0 + T) \|_1 - \| \mathbold{\Psi}(t) \|_1 \Big | \mathbold{Z}(t_0) \Big ] + c_3',
\label{driftlyapunov}
\end{aligned}
\end{equation}
where $(a)$ is true by the telescoping series, $(b)$ follows by using \eqref{evolW} to substitute $\mathbold{W}(t + 1)$, $(c)$ follows by Lemma \ref{lemma123} and the fact that the task arrival is assumed to be bounded and the service and unused service are also bounded as the number of servers are finite, so the pseudo arrival, service, and unused service are also bounded, and therefore there exists a constant $c_1$ such that $\| \mathbold{A}(t) - \mathbold{S}(t) +\widetilde{\mathbold{U}}(t)\|^2 \leq \frac{c_3'}{T}$, and $(d)$ follows by adding and subtracting the intermediary term $\langle \mathbold{W}(t), \mathbold{w} \rangle$. On the other hand,
\begin{equation*}
\begin{aligned}
& 2 \mathbb{E} \left [ \sum_{t = t_0}^{t_0 + T - 1} \Big ( \langle \mathbold{W}(t), \mathbold{A}(t) \rangle - \langle \mathbold{W}(t), \mathbold{w} \rangle \Big ) \bigg | \mathbold{Z}(t_0) \right ] \\
\leq & \ 2 \mathbb{E} \left [ \sum_{t = t_0}^{T_0 - 1} \langle \mathbold{W}(t), \mathbold{A}(t) \rangle \bigg | \mathbold{Z}(t_0) \right ] \\
& + 2 \mathbb{E} \left [ \sum_{t = T_0}^{t_0 + T - 1} \Big ( \langle \mathbold{W}(t), \mathbold{A}(t) \rangle - \langle \mathbold{W}(t), \mathbold{w} \rangle \Big ) \bigg | \mathbold{Z}(t_0) \right ] \\
\overset{(a)}{\leq} & 2 \mathbb{E} \Bigg [ \frac{(T_0 - t_0) \cdot C_A}{ \left (\min_{m, n} \{ \alpha_m^n \} \right )^2} \sum_{m \in \mathcal{M}} \Big ( Q_m^1(t_0) + \cdots + Q_m^{N^m}(t_0) \\
& + N^m \cdot C_A \cdot (T_0 - t_0) \Big ) \bigg | \mathbold{Z}(t_0) \Bigg ] \\
& + 2 \mathbb{E} \left [ \sum_{t = T_0}^{t_0 + T - 1} \Big ( \langle \mathbold{W}(t), \mathbold{A}(t) \rangle - \langle \mathbold{W}(t), \mathbold{w} \rangle \Big ) \bigg | \mathbold{Z}(t_0) \right ]
\end{aligned}
\end{equation*}
\begin{equation}
\begin{aligned}
& \hspace{-0.37cm} \leq 2 \mathbb{E} \left [ \sum_{t = T_0}^{t_0 + T - 1} \Big ( \langle \mathbold{W}(t), \mathbold{A}(t) \rangle - \langle \mathbold{W}(t), \mathbold{w} \rangle \Big ) \bigg | \mathbold{Z}(t_0) \right ] \\
& + c_2 \| \mathbold{Q}(t_0) \|_1 + c_3'',
\label{second_part}
\end{aligned}
\end{equation}
where $(a)$ is true by the fact that at most $C_A$ tasks arrive at system in each time slot, and by using the definition of pseudo task arrival in \eqref{pseudoparameters}. Putting \eqref{driftlyapunov} and \eqref{second_part} together, Lemma \ref{lemma_drift} is proved as follows:
\[
\begin{aligned}
& \mathbb{E} \Big [ V(\mathbold{Z}(t_0 + T)) - V(\mathbold{Z}(t_0)) \Big | \mathbold{Z}(t_0) \Big ] \\
\leq & \ 2 \mathbb{E} \left [ \sum_{t = T_0}^{t_0 + T - 1} \Big ( \langle \mathbold{W}(t), \mathbold{A}(t) \rangle - \langle \mathbold{W}(t), \mathbold{w} \rangle \Big ) \bigg | \mathbold{Z}(t_0) \right ] \\
& + 2 \mathbb{E} \left [ \sum_{t = t_0}^{t_0 + T - 1} \Big ( \langle \mathbold{W}(t), \mathbold{w} \rangle - \langle \mathbold{W}(t), \mathbold{S}(t) \rangle \Big ) \Big | \mathbold{Z}(t_0) \right ] \\
& + \mathbb{E} \Big [ \| \mathbold{\Psi}(t_0 + T) \|_1 - \| \mathbold{\Psi}(t) \|_1 \Big | \mathbold{Z} (t_0) \Big ] + c_2 \| \mathbold{Q}(t_0) \|_1 + c_3,
\end{aligned}
\]
where $c_3 = c_3' + c_3''$.


\subsection{Proof of Lemma \ref{asymptoticT}}
\label{proofasymptoticT}
Lemma \ref{asymptoticT}:
For any server $m \in \mathcal{M}$ and any $t_0$, we have the following:
$$\hspace{-2mm} \lim_{T \rightarrow \infty} \hspace{-1mm} \frac{ \mathbb{E} \hspace{-1mm} \left [ \sum_{t = t_0}^{t_0 + T - 1} \hspace{-1mm} \bigg ( \frac{S_m^1(t)}{\alpha_m^1} + \frac{S_m^2(t)}{\alpha_m^2} + \cdots + \frac{S_m^{N^m}(t)}{\alpha_m^{N^m}} \bigg ) \bigg | \mathbold{Z}(t_0) \right ]}{T} = 1.$$

\textbf{Proof:}
The proof is similar to the proof of lemma 6 in \cite{yekkehkhany2017gb} and is presented for the sake of completeness. Let $t_m^*$ be the first time slot after or at time slot $t_0$ at which server $m$ becomes idle, and so is available to serve another task ($t_m^*$ is also defined in \eqref{ali_t_star}); that is,
\begin{equation}
t_m^* = \min\{ \tau: \tau \geq t_0, \Psi_m(\tau) = 0 \},
\label{tm**}
\end{equation}
where, as a reminder, $\Psi_m(\tau)$ is the number of time slots that the $m$-th server has spent on the task that is receiving service from this server at time slot $\tau$.


Denote the CDF of service time of an $n$-local task on server $m$ by $F_m^n$ that has finite mean $\alpha_m^n < \infty$; therefore, $t_m^* < \infty$. 
We then have the following by considering the bounded service:
\[
\begin{aligned}
& \Bigg ( \hspace{-0.9mm} \mathbb{E} \hspace{-0.9mm} \left [ \sum_{t = t_m^*}^{t_m^* + T - 1} \hspace{-1.9mm} \bigg ( \hspace{-0.9mm} \frac{S_m^1 \hspace{-0.5mm} (t)}{\alpha_m^1} \hspace{-0.5mm} + \hspace{-0.5mm} \cdots \hspace{-0.5mm} + \hspace{-0.5mm} \frac{S_m^{N^m} \hspace{-0.7mm} (t)}{\alpha_m^{N^m}} \hspace{-0.5mm} \bigg ) \hspace{-0.5mm} \bigg | \hspace{-0.3mm} \mathbold{Z}(t_0) \hspace{-0.8mm} \right ] \hspace{-1.7mm} - \hspace{-0.7mm} \frac{t_m^* \hspace{-0.7mm} - \hspace{-0.5mm} t_0}{\alpha_m^{N^m}} \hspace{-0.5mm} + \hspace{-0.5mm} \frac{1}{\alpha_m^1} \hspace{-0.5mm} \Bigg ) \hspace{-1.35mm} \Bigg / \hspace{-1.3mm} T
\end{aligned}
\]

\begin{equation}
\begin{aligned}
& \leq \frac{ \mathbb{E} \left [ \sum_{t = t_0}^{t_0 + T - 1} \bigg ( \frac{S_m^1(t)}{\alpha_m^1} + \frac{S_m^2(t)}{\alpha_m^2} + \cdots + \frac{S_m^{N^m}(t)}{\alpha_m^{N^m}} \bigg ) \bigg | \mathbold{Z}(t_0) \right ]}{T} \leq \\
& \Bigg ( \hspace{-0.5mm} \mathbb{E} \hspace{-0.8mm} \left [ \sum_{t = t_m^*}^{t_m^* + T - 1} \hspace{-1.3mm} \bigg ( \frac{S_m^1(t)}{\alpha_m^1} + \cdots + \frac{S_m^{N^m}(t)}{\alpha_m^{N^m}} \bigg ) \bigg | \mathbold{Z}(t_0) \right ] \hspace{-0.9mm} + \frac{1}{\alpha_m^{N^m}} \Bigg ) \hspace{-0.5mm} \Bigg / \hspace{-0.5mm} T,
\label{renewalProcess}
\end{aligned}
\end{equation}
where by boundedness of $t_m^*, \alpha_m^1,$ and $\alpha_m^{N^m}$, it is obvious that $\lim_{T \rightarrow \infty}$ $\frac{- \frac{t_m^* - t_0}{\alpha_m^{N^m}} + \frac{1}{\alpha_m^1}}{T} = 0$ and $\lim_{T \rightarrow \infty} \frac{\frac{1}{\alpha_m^{N^m}}}{T} = 0$. Hence, by taking the limit of the terms in Equation \eqref{renewalProcess} as $T$ goes to infinity, we have the following:
\begin{equation}
\begin{aligned}
& \hspace{-0.29cm} \lim_{T \rightarrow \infty} \frac{ \mathbb{E} \left [ \sum_{t = t_0}^{t_0 + T - 1} \bigg ( \frac{S_m^1(t)}{\alpha_m^1} + \frac{S_m^2(t)}{\alpha_m^2} + \cdots + \frac{S_m^{N^m}(t)}{\alpha_m^{N^m}} \bigg ) \bigg | \mathbold{Z}(t_0) \right ]}{T} \\
= & \hspace{-0.03cm} \lim_{T \rightarrow \infty} \hspace{-0.03cm} \hspace{-1mm} \frac{ \mathbb{E} \left [ \sum_{t = t_m^*}^{t_m^* + T - 1} \hspace{-1mm} \bigg ( \hspace{-0.06cm} \frac{S_m^1(t)}{\alpha_m^1} + \frac{S_m^2(t)}{\alpha_m^2} + \cdots + \frac{S_m^{N^m}(t)}{\alpha_m^{N^m}} \hspace{-0.06cm} \bigg ) \hspace{-0.04cm} \bigg | \hspace{-0.02cm} \mathbold{Z}(t_0) \hspace{-0.04cm} \right ]}{T}.
\label{t_m^*}
\end{aligned}
\end{equation}
Considering the service process as a renewal process, given the scheduling decisions at the end of the renewal intervals in $[t_m^*, t_m^* + T - 1]$, all holding times for server $m$ to give service to tasks in its sub-queues are independent. We elaborate on this in the following.

We define renewal processes, $N_m^n(t), \ n \in \{1,$ $2,$ $\cdots,$ $N^m\}$, as follows, where $t$ is an integer valued number: \\
Let $H_m^n(l)$ be the holding time (service time) of the $l$-th task that is $n$-local to server $m$ after time slot $t_m^*$ receiving service from server $m$, and call $\{H_m^n(l), l \geq 1\}$ the holding process of $n$-local task type, $n \in \{1, 2, \cdots, N^m\}$.
Then define $J_m^n(l) = \sum_{i = 1}^{l} H_m^n(l)$ for $l \geq 1$, and let $J_m^n(0) = 0$. In the renewal process, $J_m^n(l)$ is the $l$-th jumping time, or the time at which the $l$-th occurrence happens, and it has the following relation with the renewal process, $N_m^n(t)$:
\[N_m^n(t) = \sum_{l = 1}^\infty \mathbb{I}_{\{ J_m^n(l) \leq t \}} = \sup \{l : J_m^n(l) \leq t \}.
\]
Another way to define $N_m^n(t)$ is as shown in the following algorithm, where by convention, $N_m^n(0) = 0$.
\begin{algorithm}
\label{algo22}
\begin{algorithmic}[1]
\STATE \emph{Set $\tau = t_m^*$, $cntr = 0$, $N_m^n(t) = 0$}

\WHILE {\emph{$cntr < t$}}
	\IF{$\eta_m(\tau) = n$}
    	\STATE $cntr++$
    	\STATE $N_m^n(t) \ += S_m^n(\tau)$
    \ENDIF
    \STATE \emph{ $\tau++$}
\ENDWHILE
\end{algorithmic}
\end{algorithm}

Another renewal process, $N_m(t)$, is defined as
\[
N_m(t) = \hspace{-0.15cm} \sum_{u = t_m^*}^{t_m^* + t - 1} \hspace{-0.15cm} \Big ( \hspace{-0.03cm} \mathbb{I}_{ \{ S_m^1(u) = 1 \}} + \mathbb{I}_{ \{ S_m^2(u) = 1 \}} + \cdots + \mathbb{I}_{ \{ S_m^{N^m}(u) = 1 \}} \hspace{-0.03cm} \Big ).
\]
Similarly, let $H_m(l)$ be the holding time (service time) of the $l$-th task after time slot $t_m^*$ receiving service from server $m$, and call $\{H_m(l), l \geq 1\}$ the holding process. Then define $J_m(l) = \sum_{i = 1}^{l} H_m(l)$ for $l \geq 1$, and let $J_m(0) = 0$. In the renewal process, $J_m(l)$ is the $l$-th jumping time, or the time at which the $l$-th occurrence happens, and it has the following relation with the renewal process, $N_m(t)$:
\[
N_m(t) = \sum_{l = 1}^\infty \mathbb{I}_{\{ J_m(l) \leq t \}} = \sup \{l : J_m(l) \leq t \}.
\]

Note that the central scheduler makes scheduling decisions for server $m$ at time slots $\{ t_m^* + J_m(l), l \geq 1 \}$. We denote these scheduling decisions by $D_m(t_m^*) = \Big ( \eta_m(t_m^* + J_m(l)) : l \geq 1 \Big )$.
Consider the time interval $[t_m^*, t_m^* + T - 1]$ when $T$ goes to infinity. Define $\rho_m^n$ as the fraction of time that server $m$ is busy giving service to tasks that are $n$-local to this server, in the mentioned interval. Obviously, $\sum_{n = 1}^{N^m} \rho_m^n = 1$. Then Equation \eqref{t_m^*} is followed by
\begin{equation}
\begin{aligned}
& \hspace{-0.1cm} \lim_{T \rightarrow \infty} \frac{ \mathbb{E} \hspace{-0.05cm} \left [ \sum_{t = t_m^*}^{t_m^* + T - 1} \hspace{-0.05cm} \bigg ( \hspace{-0.05cm} \frac{S_m^1(t)}{\alpha_m^1} + \frac{S_m^2(t)}{\alpha_m^2} + \cdots + \frac{S_m^{N^m}(t)}{\alpha_m^{N^m}} \hspace{-0.05cm} \bigg ) \bigg | \mathbold{Z}(t_0) \right ]}{T} \\
= & \lim_{T \rightarrow \infty} \Bigg \{ { \mathbb{E} \Bigg [ \mathbb{E} \Bigg [ \sum_{t = t_m^*}^{t_m^* + T - 1} \bigg ( \frac{S_m^1(t)}{\alpha_m^1} + \frac{S_m^2(t)}{\alpha_m^2} } + \\
&{ \hspace{0.15cm} \ \ \ \ \ \ \ \ \ \ \ \ \ \ \ \ \cdots + \frac{S_m^{N^m}(t)}{\alpha_m^{N^m}} \bigg ) \bigg | D_m(t_m^*), \mathbold{Z}(t_0) \Bigg ] \Bigg | \mathbold{Z}(t_0) \Bigg ]} \Bigg \} \Bigg / T \\
= & \sum_{n = 1}^{N^m} \lim_{T \rightarrow \infty} \Bigg ( \mathbb{E} \Bigg [ \frac{1}{\alpha_m^n} \mathbb{E} \Bigg [ \sum_{t = t_m^*}^{t_m^* + T - 1} \Big ( S_m^n(t)  \Big ) \\
& \ \ \ \ \ \ \ \ \ \ \ \ \ \ \ \ \ \ \ \ \ \ \ \ \ \ \ \ \ \ \ \ \ \ \ \ \ \ \ \ \ \ \bigg | D_m(t_m^*), \mathbold{Z}(t_0) \Bigg ] \Bigg | \mathbold{Z}(t_0) \Bigg ] \Bigg ) \Bigg / T\\
= & \sum_{n = 1}^{N^m} \mathbb{E} \Bigg [ \frac{1}{\alpha_m^n} \lim_{T \rightarrow \infty} \frac{\mathbb{E} \left [ N_m^n \big ( \rho_m^n T \big ) \Big | D_m(t_m^*), \mathbold{Z}(t_0) \right ]}{T} \Bigg | \mathbold{Z}(t_0) \Bigg ].
\label{t_m^*2}
\end{aligned}
\end{equation}
Note that given $\{D_m(t_m^*), \mathbold{Z}(t_0)\}$, the holding times $\{H_m^n(l), l \geq 1\}$ are independent and identically distributed with CDF $F_m^n$. If $\rho_m^n = 0$, then we do not have to worry about those tasks that are $n$-local to server $m$ since they receive service from this server for only a finite number of times in time interval $[t_m^*, t_m^* + T - 1]$ as $T \rightarrow \infty$, so
\[
\lim_{T \rightarrow \infty} \frac{\mathbb{E} \left [ N_m^n \big ( \rho_m^n T \big ) \big | D_m(t_m^*), \mathbold{Z}(t_0) \right ]}{T} = 0.
\]
But if $\rho_m^n > 0$, we can use the strong law of large numbers for renewal process $N_m^n$ to conclude the following:
\begin{equation}
\lim_{T \rightarrow \infty} \frac{\mathbb{E} \left [ N_m^n \big ( \rho_m^n T \big ) \big | D_m(t_m^*), \mathbold{Z}(t_0) \right ]}{T} = \rho_m^n \cdot \frac{1}{\mathbb{E}[H_m^n(1)]},
\label{t_m^*3}
\end{equation}
where the holding time (service time) $H_m^n(1)$ has CDF $F_m^n$ with expectation $\frac{1}{\alpha_m^n}$. Combining equations \eqref{t_m^*2} and \eqref{t_m^*3}, Lemma \ref{asymptoticT} is concluded as follows:
\begin{equation}
\begin{aligned}
& \hspace{-0.1cm} \lim_{T \rightarrow \infty} \hspace{-0.05cm} \frac{ \mathbb{E} \hspace{-0.05cm} \left [ \sum_{t = t_m^*}^{t_m^* + T - 1} \hspace{-0.05cm} \bigg ( \hspace{-0.05cm} \frac{S_m^1(t)}{\alpha_m^1} + \frac{S_m^2(t)}{\alpha_m^2} + \cdots + \frac{S_m^{N^m}(t)}{\alpha_m^{N^m}} \hspace{-0.05cm} \bigg ) \bigg | \mathbold{Z}(t_0) \right ]}{T} \\
= & \sum_{n = 1}^{N^m} \mathbb{E} \left [ \frac{1}{\alpha_m^n} \cdot \rho_m^n \cdot \alpha_m^n \bigg | \mathbold{Z}(t_0) \right ] = \sum_{n = 1}^{N^m} \rho_m^n = 1.
\end{aligned}
\end{equation}

\end{document}